\documentclass{aa}
\usepackage{graphicx,amssymb,amsmath}       
\usepackage{natbib}

\def\asca{{\it ASCA~\/}}

\def\chandra{{\it Chandra~\/}}

\def\Lsun{\hbox{$\rm L_{\odot}$}}
\def\Msun{\hbox{$\rm ~M_{\odot}$}}

\def\mbh{{M$_{\rm BH}$}}
\def\msun{{M$_{\sun}$}}

\def\H0{{\rm ~km~s^{-1}~Mpc^{-1}}}

\def\ledd{${l_{\rm Edd}}$}

\def\ergsec{{\rm ~erg~s^{-1}}}

\def\hii{\ion{H}{II}}
\def\ha{{H$\alpha$}}
\def\hanii{H$\alpha$+[N\,{\small~II}]}
\def\fullhanii{H$\alpha$+[N\,{\small~II}]~$\lambda\lambda$6548,6583}

\def\oiii{[\ion{O}{III}]}

\def\.25{0.25 keV\thinspace}

\def\d19{D$\,\leq\,$19~Mpc} 
\def\dgtr19{D$\,>\,$19~Mpc} 
\newcommand{\sgra}{Sgr~A*}

\let\lesssim=\la
\let\gtrsim=\ga

\newcommand\dummytable{\refstepcounter{table}}%

\begin{document}

\title{Radio Sources in Low-Luminosity Active Galactic Nuclei.}
\subtitle{IV. Radio Luminosity Function, Importance of Jet Power, and 
          Radio Properties of the Complete Palomar Sample}

\author{
   Neil M. Nagar 
   \inst{1} 
   \and
   Heino Falcke
   \inst{2}
   \and 
   Andrew S. Wilson
   \inst{3}
}
\offprints{Neil M. Nagar}
\institute{Kapteyn Institute, 
            Landleven 12, 9747 AD Groningen, 
	    The Netherlands \\
            Astronomy Group, Departamento de F\'isica,
            Universidad de Concepci\'on, Casilla 160-C, Concepci\'on, Chile \\
           \email{nagar@astro-udec.cl} 
           \and
           ASTRON,
	    P.O. Box 2,
	    7990 AA Dwingeloo,
	    The Netherlands \\
	    Department of Astronomy,
	    Radboud University Nijmegen,
	    Postbus 9010,
	    6500 GL Nijmegen,
	    The Netherlands \\
           \email{falcke@astron.nl}
           \and
           Department of Astronomy, University of Maryland,
             College Park, MD 20742, U.S.A. \\
           Adjunct Astronomer, Space Telescope Science Institute,
             3700 San Martin Drive, Baltimore, MD 21218, U.S.A. \\
             \email{wilson@astro.umd.edu} 
          }

\date{Received 29 October 2004; accepted 05 February 2005}

\abstract{
We present the completed results of a high resolution radio imaging survey of 
  all ($\sim$200) low-luminosity active galactic nuclei (LLAGNs) and AGNs in
  the Palomar Spectroscopic Sample of all ($\sim$488) bright northern galaxies.
The high incidences of pc-scale radio nuclei, with implied brightness 
  temperatures $\gtrsim\,10^7\,$K, and sub-parsec jets argue for
  accreting black holes in $\gtrsim$50\% of all LINERs and low-luminosity
  Seyferts; there is no evidence against \textit{all} LLAGNs being mini-AGNs.
The detected parsec-scale radio nuclei are preferentially found in massive 
  ellipticals and in type~1 nuclei (i.e. nuclei with broad \ha\ emission). 
The radio luminosity function (RLF) of Palomar Sample LLAGNs and AGNs extends three 
  orders of magnitude below, and is continuous with, that of `classical' AGNs. 
  We find marginal evidence for a low-luminosity turnover in the RLF; nevertheless 
  LLAGNs are responsible for a significant fraction of present day mass accretion.
Adopting a model of a relativistic jet from Falcke \&
  Biermann, we show that the accretion power output in LLAGNs is dominated
  by the kinetic power in the observed
  jets rather than the radiated bolometric luminosity. The Palomar LLAGNs and AGNs follow 
  the same scaling between jet kinetic power and narrow line region (NLR) luminosity as the 
  parsec to kilo-parsec jets in powerful radio galaxies.
Eddington ratios \ledd\ 
  (= L$_{\rm Emitted}$/L$_{\rm Eddington}$) of $\le$ 10$^{-1} - 10^{-5}$ are
  implied in jet models of the radio emission.
We find evidence that, in analogy to Galactic black hole candidates, LINERs are in a 
  `low/hard' state (gas poor nuclei, low Eddington ratio, ability to launch
   collimated jets) while low-luminosity Seyferts are in a `high' state (gas rich
   nuclei, higher Eddington ratio, less likely to launch collimated jets).
The radio jets are energetically more significant than 
   supernovae in the host galaxies,
   and are potentially able to deposit sufficient energy into the innermost 
   parsecs to significantly slow the gas supply to the accretion disk.
} 

\titlerunning{Radio Nuclei in Palomar Sample LLAGNs}
\authorrunning{Nagar, N. M. et al.}

\maketitle

\section{Introduction}
\label{secintro}

\subsection{Accretion activity: Ubiquity and Scaling}

It is now clear that there is no sharp division between 
``active'' galactic nuclei (AGN; i.e. nuclei presumably powered
by accretion onto a nuclear supermassive [$\gg\,10^5$ \msun] black hole) and 
``inactive'' or ``normal'' galactic nuclei (nuclei powered by star-formation-related 
processes). Rather, there is a continuous sequence of activity levels between 
these two extremes. 
There are two lines of evidence for this continuity.
The first comprises the ubiquity of black holes and the correlations between
black hole mass, galaxy bulge mass and galaxy bulge velocity dispersion
\citep{ricet98,fermer00,gebet00,merfer01,treet02,marhun03}. 
These results support the idea that many galactic nuclei are quasar relics 
\citep{sol82} and highlight the importance of studying the
coeval evolution of a galaxy and its nuclear black hole.
The second line of evidence is that many nearby galaxy nuclei not considered to be
powerful AGNs, show several characteristics in common with powerful AGNs. 
These similarities include the presence of compact radio nuclei and 
sub-parsec to 100~pc-scale radio jets \citep[e.g.][]{hec80,naget02a}, 
emission line ratios characteristic of powerful AGNs \citep[e.g.][]{hec80,hoet97a},
broad \ha\ lines \citep{hoet97b}, 
broader \ha\ lines in polarized emission than in total emission \citep{baret99},
water vapor megamasers \citep{braet97}, and
nuclear point-like UV sources \citep{maoet95,baret98}.

Important results on the growth of galaxies and their black holes, and on the 
properties and history of accretion in AGNs, are now being provided by several
large surveys, e.g. the Sloan Digital Sky Survey \citep[SDSS;][]{stoet02}.
An important complement to these large (and higher redshift) surveys of AGNs
is the study of so-called low-luminosity AGNs (LLAGNs;
i.e. low-luminosity Seyferts, LINERs, and ``transition'' 
nuclei [nuclei with spectra intermediate between those of LINERs and 
\hii\ regions]).
Here we use the term LLAGN in a more agnostic manner than AGN: we assume 
that AGNs are powered by accretion onto a supermassive black hole but make 
no {\it{a priori}} assumption about the power source of LLAGNs.
The emission-line luminosities of LLAGNs
\citep[L$_{\rm H\alpha}~\leq$ 10$^{40}$ erg s$^{-1}$ by definition;][]{hoet97a}
are a factor $\sim10^2$ times weaker than typical SDSS AGNs. If LLAGNs are truly
(weak) AGNs, then extending our studies to LLAGNs is important as they greatly 
outnumber powerful AGNs.
LLAGNs are best studied in close ($\lesssim$30~Mpc) nuclei, as a result of 
sensitivity limitations and the need to attain adequate linear resolution to 
separate any weak accretion related emission from that of the bright host galaxy.
In this paper we focus on the radio properties of the $\sim$200 LLAGNs and AGNs
(median distance $\sim$17~Mpc) in the Palomar spectroscopic sample of 
$\sim$488 bright northern galaxies \citep{hoet97a}.
The weak emission-lines of the Palomar LLAGNs
can be modeled in terms of photoionization by hot, young stars 
\citep{termel85,filter92,shi92}, 
by collisional ionization in shocks \citep{kosost76,foset78,hec80,dopsut95}
or by starbursts \citep{aloet99}.
Alternatively, they could trace AGNs accreting either at very low accretion rates
(with radiated luminosity as low as $\sim\,10^{-2}-10^{-7}$ of the Eddington 
 Luminosity, L$_{\rm Eddington}$),
or at radiative efficiencies (the ratio of radiated energy to accreted mass) 
much lower than the typical value of $\sim\,$10\% 
\citep[e.g.~Chapter~7.8 of][]{fraet95} assumed for powerful AGNs.    
 
Closely related to these theoretical and observational studies of the 
radiation from LLAGNs is the
increasing number of accurate mass determinations for 
``massive dark objects'', presumably black holes, in nearby galactic nuclei, as 
measured directly by kinematics \citep[e.g.][]{gebet03} or inferred via
the correlation between black hole mass (\mbh) and central stellar velocity dispersion
\citep[$\sigma_c$;][]{fermer00,gebet00,merfer01,treet02} or galaxy bulge mass 
\citep{ricet98,marhun03}. 
These mass determinations, coupled with the emitted luminosity from the AGN, 
enable a measure of the Eddington ratio, i.e. 
the emitted accretion luminosity in units of the
Eddington luminosity (\ledd$\,=$ L$_{\rm Emitted}$/L$_{\rm Eddington}$). 
In this paper we argue that accounting for the kinetic power in the
radio jet is crucial when estimating L$_{\rm Emitted}$ (and hence \ledd) 
in LLAGNs even 
though the radiated luminosity in the radio band is bolometrically unimportant.
Our high resolution radio observations of a large number of nearby LLAGNs 
considerably increase the number of LLAGNs with reliable black hole mass
estimates \textit{and} high resolution radio observations, allowing 
a better test of the relationship between these quantities. 

\subsection{Identifying weak AGNs via their Radio Emission}

As discussed in Sect.~1 of \citet{naget02a}, a sub-parsec, high brightness
temperature (T$_{\rm b}\,\gtrsim\,10^7\,$K), flat spectrum nuclear radio source
and any radio jets are reliable 
indicators of the presence of an accreting supermassive black hole.
Interestingly, compact flat spectrum radio nuclei are also detected toward the 
$\sim$10-15 \msun\ black holes in Galactic X-ray binary sources, especially during 
phases of highly sub-Eddington accretion \citep[see review by][]{fenbel04}.
The only known sources of log~[T$_{\rm b}~$(K)]$\,\geq$ 5 radio emission
in compact starbursts are radio supernovae (RSNe) within the starburst 
\citep[e.g.][]{conet91,smiet98}. However, even RSNe or groups of RSNe cannot 
reproduce the compactness, high brightness temperatures, and flat spectral 
indices seen in the radio nuclei of LLAGNs (discussed in Sect.~\ref{secdiscussion}).
It then only remains to use the radio morphology and spectral shape to test whether 
the radio emission originates from a jet (possibly relativistic) launched by the 
black hole or from the accretion flow itself. Radio emission is expected from
radiatively inefficient accretion flows (RIAFs), e.g. advection-dominated 
\citep[ADAF;][]{naret98} or convection-dominated \citep[CDAF;][]{naret00} 
accretion flows, possible forms of accretion onto a black hole at low accretion 
rates \citep{reeet82}. 

We have argued \citep{naget02a} that the combination of the 
Very Large Array{\footnote{The VLA and VLBA are operated by
the National Radio Astronomy Observatory, a facility of the National Science
Foundation operated under cooperative agreement by Associated Universities, Inc.}}
\citep[VLA;][]{thoet80} and the Very Long Baseline Array$^{1}$ \citep[VLBA;][]{napet94}
makes a highly effective and efficient tool to unambiguously identify weak AGNs in 
bright galactic nuclei.
The main advantages are the minimal obscuration at high gigahertz frequencies,
the high resolution which allows one to easily pick out the AGN as most 
radio emission from other sources is usually resolved out, and the high sensitivity.

\subsection{Previous and Future Work in this Series}

The effectiveness of radio searches for AGNs in LLAGNs is borne out by the
results of previous papers in this series. These include results of VLA and 
VLBA observations of 48 LINERs \citep[][Papers~I and II]{naget00,falet00}, and a 
distance-limited (\d19) sample of 96 LLAGNs \citep[][Paper~III]{naget02a}, from the
Palomar Sample. These papers showed that $\gtrsim$50\% of all LINERs and low-luminosity 
Seyferts have compact flat-spectrum radio nuclei at 150~mas resolution. Follow-up VLBA
imaging attained a 100\% detection rate of high brightness temperature milli-arcsec
scale nuclei in a radio-flux limited subsample. The compactness, 
high brightness-temperature ($\geq10^6$~K), and
other properties, all argue for an origin of the radio emission in AGN-related processes.
The morphology (sub-parsec jets are detected in several nuclei) and radio spectral shape
\citep{naget01,naget02b} support the dominant source of radio emission as the self-absorbed
base of a relativistic jet launched by the black hole, rather than a radiatively
inefficient accretion inflow.
Compact radio nuclei are preferentially found in massive ellipticals and in type~1 nuclei.
The core radio luminosity is correlated with the nuclear optical `broad' \ha\ luminosity, 
the nuclear optical `narrow' emission-line luminosity and width, and the galaxy 
luminosity \citep{naget02a}. In these correlations, LLAGNs fall close to the low-luminosity
extrapolations of more powerful AGNs. 
The sub-arcsec radio luminosity is correlated with both the estimated mass of the 
nuclear black hole and the galaxy bulge luminosity.
Partial correlation analysis on the two correlations yields the result that
each correlation is meaningful even after removing the effect of the other
correlation \citep{naget02a}.

This paper presents completed results of our high resolution radio imaging survey 
of all LLAGNs and AGNs in the Palomar Sample.
Future papers will present results on the 1.4~GHz to 667~GHz radio spectral shapes
of a subsample of LLAGNs 
\citep[Nagar et al., in prep; preliminary results in][]{naget02b} and the sub-pc jet 
morphology and jet proper motions in LLAGNs (Nagar et al., in prep).

\subsection{Organization of this Paper}

In the following sections, we first define the sample used 
(Sect.~\ref{secsample}) and then summarize previous observations and 
report on new VLA and VLBA observations which complete the radio survey of 
LLAGNs and AGNs in the Palomar sample (Sect.~\ref{secobspal}).
The results of all the VLA and VLBA radio observations of the Palomar 
sample are presented in Sect.~\ref{secres}.
These results are used to synthesize an overall picture of the incidence and 
properties of AGNs in LLAGNs -- including the radio luminosity function, 
importance of jet energetics, and correlations with other emission-line and host 
galaxy properties -- and their continuity with more powerful AGNs 
(Sect~\ref{secderived}). The results are briefly discussed in Sect.~\ref{secdiscussion}
and the major conclusions of the completed radio study are listed in 
Sect.~\ref{secconclusion}. Finally, the appendix contains a compilation of high 
resolution radio observations of the 53 absorption-line nuclei in the Palomar sample.
In this paper, as in previous papers of this series, we use a Hubble constant
H$_0\,= 75$ km s$^{-1}$ Mpc$^{-1}$ to be consistent with \citet{hoet97a} who
tabulate the results of optical spectroscopy of the Palomar sample.
In this paper we use `radio luminosity' to denote the radiated power at a given
radio frequency and `jet power' or `jet kinetic power' to denote the kinetic
or mechanical power in the jet (as derived from models of relativistic jets).

\section{Sample}
\label{secsample}

The results in this paper, and the new observations reported here, are
based on LLAGNs and AGNs selected from the Palomar spectroscopic survey of all
($\sim\,$488) northern galaxies with B$_{\rm T} <$~12.5~mag \citep{hoet95}. 
Spectroscopic parameters (including activity classification) of 418
galaxies in the Palomar spectroscopic survey which show nuclear emission lines 
have been presented in \citet{hoet97a}; updates to these,
and upper limits to the
emission-line fluxes of a further 53 nuclei without detected emission-lines,
are presented in \citet{hoet03a}. 
Of the 418 galaxies
with nuclear emission lines in \citet{hoet97a}, we consider only the 403
which belong to the defined Palomar sample.
We thus included in our radio survey approximately 7 AGNs and
190 LLAGNs (using the operational boundary of
L$_{\rm H\alpha}~\leq$ 10$^{40}$ erg s$^{-1}$
to distinguish LLAGNs from AGNs: Ho et al. 1997a). The 206 nuclei with
\hii\ region type spectra which make up the balance of the 403 are excluded
from our survey.
Of the 7 AGNs, only two (NGC~1275 and NGC~4151) have \ha\ luminosities 
significantly greater than the boundary between AGNs and LLAGNs. The other
five AGNs are within a factor $\sim$2 of the boundary. In view of the 
significant photometric uncertainties (for some of these nuclei) and the large 
aperture of the H$\alpha$ luminosity measurements, these 5 AGNs could be loosely
considered as LLAGNs. Thus, while we 
use `Palomar LLAGNs and AGNs' to describe the sample observed in the
radio, it is worth bearing in mind that this sample is almost exclusively 
comprised of LLAGNs.
We have not observed the 53
nuclei without detected emission lines \citep{hoet03a}, but
list their radio flux densities, derived from the literature, in the Appendix.

\section{Radio Observations of the Palomar Sample}
\label{secobspal}

Several earlier surveys have targeted a substantial fraction of the 
nearby galaxies which are now in the Palomar Sample.
Radio surveys with resolution $\sim$1{\arcsec}-10{\arcsec} include 
those by \citet{hecet80,hum80,wrohee84,humet87,fabet89,caret90,wrohee91,lauet97} 
and \citet{wroet04}.
Higher resolution surveys (VLBA or VLBI) include those by 
\citet{jonet81} and \citet{humet82}.

Since the publishing of comprehensive optical results on the Palomar Sample,
three groups have conducted large radio surveys of the sample. 
Our group has now completed a 0{\farcs}15 resolution 15~GHz (2~cm) VLA survey
of all LLAGNs except some transition nuclei at \dgtr19 
\citep[a total of 162 nuclei observed;][this work]{naget00,naget02a}. 
We then observed all strong sources with the VLBA unless they had already
been observed at VLBI resolution \citep[][this work]{falet00,naget02a}.
\citet{houlv01} and \citet{ulvho01a} have observed all (45) 
Palomar Seyferts at arcsec resolution
at 5~GHz (6~cm) and 1.4~GHz (20~cm) and followed up the strong detections at
multiple frequencies with the VLBA \citep{andet04}.
\citet{filet00} and \citet{filet04} have completed a 5{\arcsec}-0{\farcs}3 
resolution survey of all transition nuclei in the sample with follow up 
VLBA observations of some of the stronger nuclei.
Finally, \citet{ulvho01b} have completed a survey of a well-defined
sub-sample of 40 \hii\ type nuclei in the Palomar sample and found that 
none of them has a compact radio nucleus at the flux levels of those in 
LLAGNs in the sample. The latter result justifies the exclusion of 
\hii\ type nuclei from the remaining discussion of this paper.

With the new VLA observations reported here,
all except 4 of the LLAGNs and AGNs in the Palomar sample have now 
been observed at sub-arcsec resolution with the VLA.
The four exceptions are:
NGC~5850, 
NGC~5970, 
NGC~5982, and 
NGC~5985. 
We believe that none of these four nuclei would have been detected in our survey,
since their measured fluxes are $<$1~mJy in observations 
at 1.4--5~GHz with 1{\arcsec}--5{\arcsec} resolution \citep{humet87,wrohee91}.
With the new VLBA observations reported here,
all Palomar LLAGNs and AGNs with S$^{\rm VLA}_{\rm 15\,GHz} >$ 2.7~mJy
(except NGC~5377) have been observed at milli-arcsecond resolution with the VLBA.

\subsection{New VLA Observations and Data Reduction}
\label{secobsvla}

Fifty one Seyferts and LINERs at \dgtr19 were observed at 15~GHz (2~cm) 
with the VLA in a 14~hr run on 2001 January 13 and 14.
The VLA was in A-configuration \citep{thoet80} at this time 
and was configured to observe in full polarization mode with two channels 
(``IF''s) of 50~MHz each.
Most target sources were observed at elevations between
40{\degr} and 60{\degr}; only a few southern sources were
observed at lower elevations, but always above 33{\degr}. 
Each target source was observed with a 7~min
integration sandwiched between two 1~min observations
on a nearby strong point-like source (the `phase calibrator').
Typically, each target galaxy was observed once in this way;
for a handful of galaxies, we were able to make two such passes.
The following galaxies not in the Palomar sample, but selected for having accurate
black hole mass measurements \citep{ricet98}, 
were also observed during the run:
NGC~205, NGC~221, NGC~821, NGC~1023, NGC~2300,
NGC~7332, NGC~7457, and NGC~7768. 

Data were calibrated and mapped using AIPS, following the 
standard reduction procedures as outlined in the AIPS 
cookbook\footnote{available online at www.nrao.edu}. 
Elevation dependent effects were removed using the post-October
2000 antenna gain solutions, along
with corrections for the sky opacity during the run.
Observations of 3C~286 (observed at elevation 50{\degr}) 
were used to set the flux-density scale at 15~GHz. A second flux
calibrator, 0410+769 (a.k.a. 0409+768; observed at elevation 47{\degr}), 
was also observed as a flux check source.
The 1$\sigma$ error in flux bootstrapping (i.e. setting the flux density 
scale relative to the flux calibrators) is expected to be roughly 2.5\%. 

Maps in Stokes I were made with task IMAGR.
Since most targets were observed in `snapshot mode' (i.e. for a 
short period at a single hour angle), the synthesized beam was not optimal, 
some of the maps have a low signal-to-noise ratio and any extended structure
would not have been mapped properly.
For sources stronger than about 3~mJy, we were able to iteratively 
self-calibrate the data so as to increase the signal-to-noise ratio in the final 
map.  The total flux of the source did not change appreciably during this self
calibration process; we therefore did not scale up (to compensate for errors in 
the phase calibration process) the fluxes of sources weaker than 3~mJy.
The resolution of the final maps was $\sim\,$0{\farcs}15.
The typical root mean square (r.m.s.) noise in the final maps was 0.3~mJy
and we use a formal detection limit of 1.5~mJy (i.e. 5$\,\sigma$).
However, weaker sources (down to 0.7~mJy) have been tentatively
detected in some nuclei with positions coincident with the highly accurate 
positions of the optical nuclei (Cotton, Condon, \& Arbizzani 1999).

\subsection{New VLBA Observations and Data Reduction}
\label{secobsvlba}

In order to obtain uniform mas-resolution maps of \textit{all} 
LLAGNs and AGNs in the Palomar sample with 
S$^{\rm VLA}_{\rm 15\,GHz} >$ 2.7~mJy, we selected 10 LLAGNs from 
Table~\ref{tabvla} which had not previously been observed at high 
enough resolution, signal-to-noise, or image fidelity with the VLBA or 
VLBI at 5~GHz.
These ten LLAGNs, plus one galaxy with an \hii\ type nucleus (NGC~3690),
were observed with the VLBA at 4.9 GHz (6~cm) in a 15 hour run on 
December 17, 2001.

All observations were performed in single polarization (`LL') mode, with
128 Megabits per second bitrate, and with 4 channels (``BB''s) of bandwidth
8~MHz each.
Each source was observed at two or three different hour angles in order to 
obtain good $(u,v)$-coverage.  For each source, the first observation pass used a 
cycle of 4~min on source and 1 min on a nearby 
(distance 1.5\degr--5.5\degr)
phase calibrator, repeated seven times. 
The second pass used a cycle of 2~min on source and 1~min on a 
nearby phase calibrator, repeated 13 times. 
The total integration time on each source was therefore at least 
(and typically) 54~min. 
The ``fringe finder'' sources J0555+3948 and J0927+3902 were briefly 
observed and later used for first order synchronization of the 
data from the different antennas.
The weather was mostly fair at all VLBA sites. 
There were a few
intermittent tape problems at several antennas; all data points with tape 
weights less than 0.7 (on a scale of 0 to 1) were deleted.
Data were calibrated using AIPS, closely following the procedures 
in the `VLBA pipeline' \citep{sjoet04}.
Bad $(u,v)$ data were deleted before the phase solutions of the 
phase-calibrator observations were transfered to the galaxy data.

Images in Stokes I (assuming no circular polarization) of the sources 
were made using AIPS task IMAGR.
For sources stronger than about 3~mJy, we were able to iteratively
self-calibrate and image the data so as to increase the signal-to-noise 
in the final map. 
The peak flux-density of the source typically increased by a factor of 1.3 
during the self-calibration process. Therefore, for sources
weaker than 4 mJy, on which accurate self-calibration was not possible,
we have multiplied the {\it{peak}} detected flux-density by 1.3 as a crude
attempt to correct for atmospheric decorrelation losses. The r.m.s.
noise in the final, uniformly-weighted images is typically 0.15~mJy to 
0.2~mJy, and the resolution between 2~mas and 5~mas.

\section{Results of the Radio Observations}
\label{secres}

\subsection{Results of the VLA Observations}
\label{secvlares}

\begin{figure}[ht]
\resizebox{\columnwidth}{!}{
  \includegraphics[bb=75 230 540 390,clip,width=3.6in]{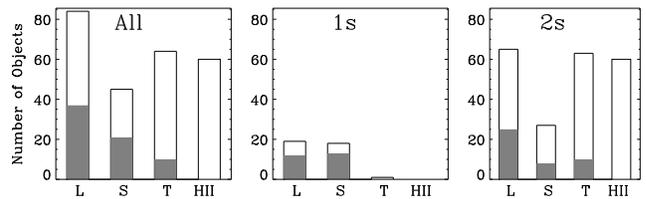}
}
\caption{Detection rate of 15~GHz 150-mas-scale radio nuclei for 
``L''INERs, ``S''eyferts, and ``T''ransition nuclei in the Palomar sample.
The total number of objects is shown by the upper histogram and the number detected
is shown by the grey-shaded histogram.
Note the higher detection rates of type~1 (i.e. galaxies with broad \ha\ emission)
Seyfert and LINER nuclei.
}
\label{figdetrate}
\end{figure}

The detection rate of radio nuclei with the VLA is illustrated in 
Fig.~\ref{figdetrate}.
When all LLAGNs and AGNs in the Palomar 
sample are considered, the VLA observations have detected 
21 of 45 (47\%$\pm$10\%) Seyferts,
37 of 84 (44\%$\pm$7\%) LINERs, and 
10 of 64 (16\%$\pm$5\%) transition nuclei, at a resolution
of $\sim\,$0{\farcs}15 and above a flux limit of 1--1.5~mJy. Additionally,
one \hii\ type nuclei (NGC~3690) was also detected.
Alternatively, one can state that radio nuclei
with luminosity L$^{\rm core}_{\rm 15\,GHz}~\geq$ 10$^{20}$ W~Hz$^{-1}$
are found in 
15 of 45 Seyferts, 
27 of 84 LINERs, and 
 6 of 64 transition nuclei.
The radio luminosities of the detected 15~GHz nuclei lie between
10$^{18}$ and 10$^{23}$ W Hz$^{-1}$, similar to the
luminosities seen in `normal' E/S0 galaxies \citep{sadet89}. 
It is notable, however, that a significant fraction of
the detected 15~GHz compact nuclei are in spiral galaxies.
Most of the detected 15~GHz nuclear radio sources are compact at the 
0{\farcs}15 resolution (typically 15--25~pc) of our survey: the implied 
brightness temperatures are typically 
T$_b$~$\geq$ 10$^{2.5-4.0}$~K.

A complete list of results of the VLA observations of all Palomar LLAGNs and 
AGNs
appears in Table~\ref{tabvla} with columns explained in the footnotes. 
All table columns are listed for sources observed by us;
for data taken from the literature we list only the peak and total flux, 
the peak luminosity and the reference in which these and the remaining 
data of the table can be found.
For the new VLA observations reported here (Sect.~\ref{secobsvla}) the 
radio positions of detected nuclei were measured directly in J2000 coordinates. 
For detections reported earlier in \citet{naget00} and \citet{naget02a}, we 
have precessed the B1950 coordinates reported there to equinox J2000 using 
the SCHED software (the software used to create VLBA observing scripts).
The radio positions for the detected nuclei are limited by the positional 
accuracy of the phase calibrators (typically 2--10~mas), 
and by the accuracy of the Gaussian fit to the source brightness distribution, 
which depends on the signal-to-noise ratio of the source detection. 
The overall accuracy should typically be better than $\sim\,$50~mas.
We have compared the radio positions derived here with optical positions 
from \citet{cotet99a}, which were measured from the digital
sky survey with typical 1$\sigma$ accuracy 1{\farcs}5--2{\farcs}5 in
each of right ascension and declination. The results (col. 7 of 
Table~\ref{tabvla}) show a good ($\leq$~2$\sigma$) agreement in most cases.
Very few nuclei show reliable extended structure in our 15~GHz
maps; the absence of extended emission in most nuclei
is not surprising as the high resolution may resolve it out. In addition, 
such extended emission is expected to be weak at the high frequency observed.

For all the 8 additional sources (i.e. not in the Palomar sample)
with accurate black hole masses and 
observed in the January 2001 VLA run (Sect.~\ref{secobsvla}), we can place 
firm 5$\sigma$ upperlimits of 1.5~mJy on the nuclear radio emission.

\begin{figure*}[ht]
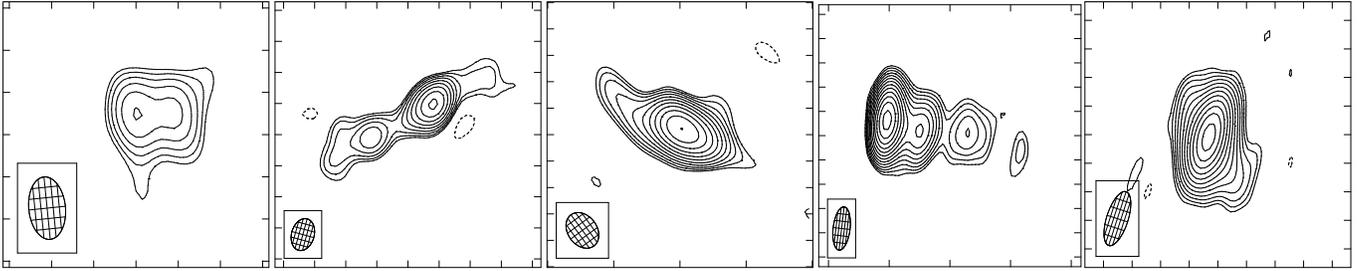

\resizebox{\textwidth}{!}{
  \includegraphics[bb=75 220 560 700,clip]{n2273.ps}
  \includegraphics[bb=75 220 560 700,clip]{n4589.ps}
  \includegraphics[bb=75 220 560 700,clip]{n5353.ps}  
  \includegraphics[bb=82 222 556 697,clip]{n5363.ps}
  \includegraphics[bb=75 220 560 700,clip]{n7626.ps}}
\caption{
 5~GHz (6~cm) VLBA maps of (left to right) NGC~2273, NGC~4589, NGC~5353, 
 NGC~5363, and NGC~7626.
 The contours are integer powers of $\sqrt{2}$, multiplied by the 
 $2\,\sigma$ noise level of 0.25~mJy (0.3~mJy in the case of NGC~5353).
 The peak flux-densities are 20.6, 6.1, 13.7, 26.1, and 19.0~mJy/beam, respectively.
 Tick marks are every 2~mas except for NGC~5353 and NGC~5363 (5~mas).
}
\label{figvlba}
\end{figure*}

\subsection{Results of the VLBA or VLBI Observations}
\label{secvlbares}

All ten LLAGNs newly observed with the VLBA at 5~GHz were 
clearly detected in initial maps (i.e. without any form of self-calibration).
The single \hii\ type nucleus observed, NGC~3690, was not detected in our 
observations; a weak (1.5~mJy) high brightness-temperature ($>10^7$~K) radio 
nucleus in this source was detected in previous deep 1.4~GHz VLBI observations 
\citep{lonet93}.
Of the nuclei detected in our observations, mas-scale radio cores were already 
known to exist in
NGC~1167 \citep{sanet95,gioet01}, 
NGC~2911 \citep{schet83,sleet94,filet02},
NGC~5353, NGC~5363 \citep{humet82}, 
and NGC~7626 \citep{xuet00}.
Additionally, NGC~2273  \citep{lonet92} was
also suspected of having a mas-scale radio core. Contour maps of
the sources found to be extended in our new VLBA observations are shown
in Fig.~\ref{figvlba}. Images of the other radio nuclei with extended
mas-scale structure can be found in 
\citet{falet00}, \citet[][and references therein]{naget02a}, \citet{andet04}, 
and \citet{filet04}. 

The VLBA observations confirm that all except one (NGC~2655) nucleus with
S$_{\rm VLA}^{\rm 15\,GHz}\,>$ 2.7~mJy are genuine AGNs with the radio 
emission coming from mas- or sub-parsec-scales.
A compilation of the results of all VLBA and VLBI observations of LLAGNs 
and AGNs in the Palomar sample is presented in Table~\ref{tabvlba}, 
with columns explained in the footnotes. Of the 44 sources listed in the
table, 39 are from the flux-limited sample (i.e.  S$^{\rm VLA}_{\rm 15\,GHz} >$ 2.7~mJy
in Table~\ref{tabvla}).
Radio positions of sources with data taken from the literature can be found 
in the references listed in col. 18.
For sources from this and our previous works we list source positions
referenced to the positions of their respective phase-calibrators.  
All of these source positions have been updated to reflect the latest 
(as of January 2004) phase calibrator positions 
\citep[expected accuracy $\sim$1~mas;][]{beaet02}.
The other factors contributing to the
position uncertainty are the accuracy of the Gaussian fit to the 
target source - which should typically be better than 2~mas - and 
the error in transfering the phase-calibrator position to the source -
which is expected to be better than 3~mas given the small angular
separations between our source-phase-calibrator pairs
\citep[see e.g. Fig. 3 of][]{chaet04}.
Thus the overall accuracy of the positions listed in Table~\ref{tabvlba} 
should be better than 5~mas.
The implied brightness 
temperatures were calculated using the formula given in \citet{falet00};
the results are in the range 
$>$10$^{6.3}$K to $\gtrsim10^{10.8}$K. 
Since most of the sources are unresolved, these values are lower 
limits to the true brightness temperatures.

It is interesting to note that the VLBI-detected nuclei in Table~\ref{tabvlba} 
include two galaxy pairs in which both members of the pair host an AGN. 
NGC~3226 and NGC~3227 (which together form Arp~94) are a galaxy pair 
(inter-nuclear distance $\sim\,$2{\farcm}2 or $\sim\,$13~kpc)
in a strong encounter with prominent tidal plumes.
NGC~5353 and NGC~5354 have a nuclear separation of $\sim$1{\farcm}2 or $\sim$12~kpc, 
and are members of Hickson compact group 68 (HCG~68). 

\section{Radio Properties}
\label{secderived}

\subsection{Correlations between Radio Luminosity, Optical Emission-Line and
         Host Galaxy Properties}
\label{secemi}

\begin{figure*}
\resizebox{\textwidth}{!}{
\includegraphics[bb=72 216 504 446,clip]{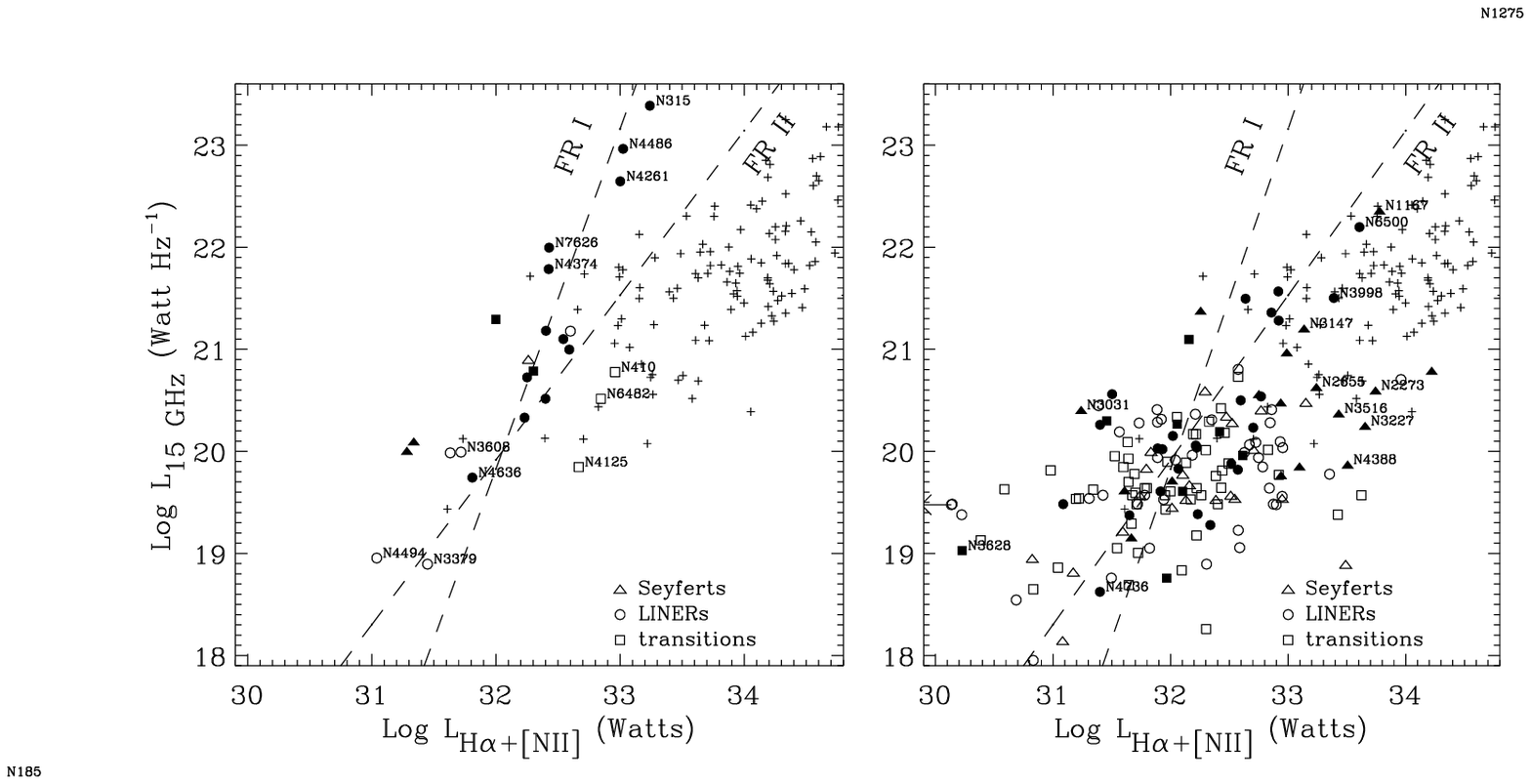}}
\caption{
 A plot of the log of the 15~GHz nuclear (150~mas resolution) radio luminosity versus
 nuclear \fullhanii\ luminosity for all LLAGNs and AGNs in the Palomar sample
 in elliptical (left) and non-elliptical (right) host galaxies.
 In both panels, Seyferts, LINERs, and transition nuclei from the Palomar sample
 are shown as triangles, circles, and squares, respectively. For these,
 filled symbols are used for radio detected nuclei and open symbols 
 are used for upper limits to the radio luminosity.
 For comparison, radio-detected
 ``classical'' Seyfert galaxies (from Whittle 1992a, i.e. not from the Palomar Sample) 
 are plotted as crosses in both panels, regardless of their galaxy morphological type 
 (for these the \oiii\ luminosity was converted to an \hanii\
 luminosity using standard flux ratios for Seyferts; see Nagar et al. 2002a).
 Also shown are the low-luminosity extrapolations of linear fits to the same
 relationship for FR~I and FR~II radio galaxies \citep[dashed lines;][]{zirbau95}.}
\label{figradvshan2}
\end{figure*}

Correlations between radio, optical emission line, and host galaxy properties in the 
Palomar Sample have been addressed in detail for
all Seyferts \citep{ulvho01a}, 
all (96) LLAGNs at \d19 \citep{naget02a}, and 
all transition nuclei \citep{filet04}.
The results from these papers are essentially unchanged after expanding the sample
to include all LLAGNs and AGNs in the Palomar sample.
Here we present only an update on the relation between radio luminosity and optical 
emission-line luminosity (Fig.~\ref{figradvshan2}).
The elliptical radio detections (filled symbols in the left panel) are closely 
related to FR~Is, as found earlier by \citet{naget02a}. 
The late-type Palomar Seyferts detected in the radio (filled triangles in the 
right panel) on the other 
hand lie closer to the region (and its low luminosity extrapolation) occupied
by `classical' Seyferts \citep[see also][]{naget02a}. 

\subsection{Radio Luminosity and Black Hole Mass}

\begin{figure*}[ht]
\resizebox{\textwidth}{!}{
\includegraphics{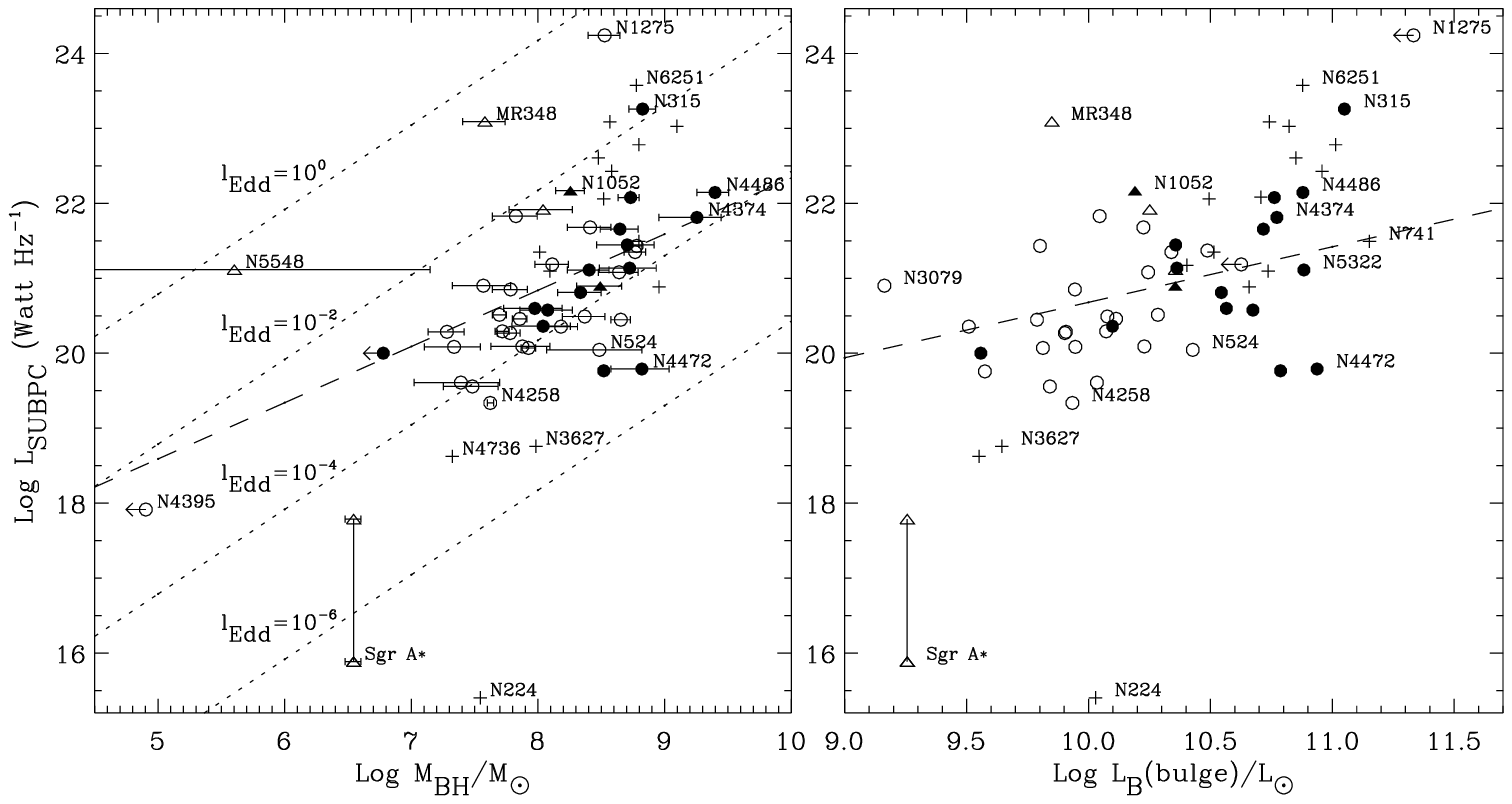}}
\caption{
 A plot of sub-parsec radio luminosity versus
 (left) black hole mass and 
 (right) bulge luminosity of the host galaxy in the B-band. 
 Only radio-detected sources relatively unambiguously identified with
 the central engine of the AGN (see text) and with radio fluxes measured
 at resolution better than 1~pc are plotted as circles (Palomar LLAGNs and
 AGNs) and triangles (other LLAGNs and AGNs).
 For these, filled symbols are used for elliptical galaxies,
 and errors in black hole mass are shown (see text). 
 LLAGNs and AGNs (some of which are in the Palomar sample) 
 with radio luminosities measured at resolution between 
 1~pc and 5~pc are shown as crosses.
 Two measurements \citep[at different resolutions;][]{naget02a} are plotted 
 for the Galaxy. 
 The four dotted lines in the left panel represent
 Eddington ratios (from 10$^{-6}$ to 1) calculated assuming that jet kinetic power 
 dominates the accretion energy output (see Sect.~\ref{secjetpower}).
 The dashed line in the left panel shows a linear fit to the circles and
  triangles with log \mbh $>$ 10$^7$ \msun; the dashed line in right panel shows
  a linear fit to the circles and triangles except the Galaxy (see text). 
 }
\label{figvlbamdo}
\end{figure*}

Correlations between radio luminosity, black hole mass, and galaxy luminosity
have been discussed in \citet{naget02a} using all of the
150~mas resolution VLA radio data listed in this paper along with
sub-arcsecond resolution radio data on other nearby galaxies with
black hole mass (\mbh) estimates. \citet{naget02a} found that the radio luminosity is 
correlated with both the black hole mass and the bulge luminosity at the 99.99\% 
significance level. Partial correlation analysis on the two correlations 
yielded the result that each correlation is meaningful even after removing 
the effect of the other correlation.
Since then a `fundamental plane' between black hole mass, X-ray luminosity,
and radio luminosity, which fits both Galactic black hole candidates and AGNs
has been claimed by \citet{meret03} and \citet{falet04}.

Here we refine the correlations presented in \citet{naget02a} by considering 
only nuclei observed with linear resolution $\leq\,1\,$pc in the radio and for 
which one radio component can be relatively unambiguously identified with the 
location of the central engine. This resolution and morphological criterion 
enables a more accurate measure of the radio emission from only the 
accretion inflow and/or the sub-parsec base of the jet, and helps avoid 
contamination from radio emission originating in knots further out in the jet. 
The latter radio emission is common in LLAGNs (Sect.~\ref{secjets}) and 
often dominates the parsec scale radio emission in Seyferts. In fact 
many Seyferts have several radio sources in the inner parsec, none of which
are unambiguously identifiable with the central engine
\citep{kuket99,munet03,midet04}.

Fig.~\ref{figvlbamdo} shows the relation between sub-parsec radio luminosity, 
\mbh, and host galaxy bulge luminosity in the B-band. Black hole
masses are either directly measured from stellar, gas, or maser
dynamics, or estimated from the central stellar velocity dispersion, $\sigma_c$ 
\citep[][for all other references see Nagar et al. 2002a]{emset99,gebet03}.
We used the relationship of \citet{treet02} to estimate \mbh\ from $\sigma_c$.
For the circles and triangles in the plot, we also show
the error in the black hole mass determination (Fig.~\ref{figvlbamdo}a). These errors
represent reported 1$\sigma$ errors for masses measured directly from stellar, gas, 
or maser dynamics \citep[e.g.][]{gebet03}, or reflect 1$\sigma$ errors in the reported 
central stellar velocity dispersion ($\sigma_c$) assuming no additional error in 
converting $\sigma_c$ to black hole mass \citep{treet02}.

The plotted circles show the 44 galaxies in Table~\ref{tabvlba}, except
NGC~266 (linear resolution 1.3~pc; plotted as a cross),
NGC~1167, NGC~4772 (no measurement of $\sigma_c$ in the literature), 
and NGC~2655 (not detected with the VLBA).
NGC~4395 
\citep[in the Palomar sample, and detected in deep VLBA 
observations;][]{wroet01} is taken to not have a bulge 
\citep{filho03}. 
In addition, we plot (as triangles) 6 galaxies which are not
Palomar LLAGNs or Palomar AGNs and which have 
radio nuclei relatively unambiguously identified with
the central engine in maps with resolution better than 1~pc, and
available black hole mass measurements or estimates from $\sigma_c$.
These 6 galaxies are: 
the Galaxy \citep{kriet98}, for which as in \citet{naget02a} we plot two 
radio luminosities: that of only \sgra\ (10~mas or 4$\times10^{-4}$ pc resolution) and 
that for the full Sgr complex; and 
Seyferts observed with the VLBA/I \citep[from the list compiled in][]{midet04}:
Mrk~348 \citep{ulvet99}, 
NGC~1052 \citep{kelet98},
NGC~2110, NGC~5252 \citep{munet00},
and NGC~5548 \citep{wro00}.

\citet{xuet00} observed additional (i.e. not in the Palomar Sample) FR~Is 
with the VLBA, but their resolution of $\sim$7~mas translates
to $>$1~pc and only some of these FR~Is could be included in the figure as crosses.
We do not consider other more powerful radio sources (e.g. blazars) to
minimize confusion due to relativistic beaming.

A visual inspection of Fig.~\ref{figvlbamdo} shows a rough overall correlation between
radio luminosity and both black hole mass and galaxy bulge luminosity.
The large scatter in both relationships is possibly due to a large range of accretion 
rates at any given black hole mass (dotted lines in Fig.~\ref{figvlbamdo} left), 
which results in widely different output radio luminosities.
Statistical tests{\footnote{Measurement errors were 
not considered when running the statistical tests.}} from the ASURV package 
\citep{lavet92}, suggest that both correlations are statistically significant 
even when the Galaxy and other nuclei with low black hole mass are removed, 
as detailed below.
The radio luminosity and black hole mass correlation has significance 99.95-99.98\% when
all circles and triangles in Fig.~\ref{figvlbamdo} (left panel) are considered.
This significance drops only slightly (98.9-99.8\%) when the Galaxy and NGC~4395 
are not considered, and is still 98.9-99.8\% when only nuclei with log \mbh
$>$ 10$^7$ \msun\ are considered.
The correlation between radio luminosity and bulge luminosity has significance 
99.3-99.7\% when all circles and triangles in Fig.~\ref{figvlbamdo} (right panel) are 
considered. This correlation is still significant (95.6-99.2\%) when the Galaxy is not
considered.
Linear regression analysis by the Buckley James method in ASURV on the circles and 
triangles in Fig.~\ref{figvlbamdo}
 (not considering the Galaxy in both cases, and using only 
nuclei with log \mbh $>$ 10$^7$ \msun\ in the case of the relation between radio 
luminosity and \mbh) yields:
\begin{equation}
\begin{split}
\noindent {\rm log\,(L}_{\rm Sub-pc}\,[{\rm W/Hz}])\,&    \\
     & \hspace{-0.8in} = 0.74\hspace{0.04in}(\pm0.31)\; {\rm log}\,({\rm L}_{\rm B}({\rm bulge})/{\rm L_\odot}) + 13.28 \\
     & \hspace{-0.8in} = 0.75\hspace{0.04in}(\pm0.26)\; {\rm log}\,({\rm M}_{\rm BH}/{\rm M_\odot}) \hspace{0.27in} + 14.84  \\
\end{split}
\end{equation}
These two relations are plotted with dashed lines in Fig.~\ref{figvlbamdo}.

The relation 
between radio luminosity and black hole mass has been discussed 
by several authors, with conflicting results. 
For example, \citet{fraet98} claimed a correlation with a much steeper slope
based on a small number of objects, while \citet{ho02} and
\citet{woourr02} found no correlation for larger 
samples of AGNs and LLAGNs. We emphasize that our results and those of \citet{ho02}
and \citet{woourr02} are not contradictory given the samples
and physical scales of the radio emission.
Here, in Fig. 4, we specifically address the correlation
between radio emission from the base of the
jet or from the innermost accretion inflow 
(measured here by the sub-parsec radio emission in nuclei for which
this radio emission is relatively unambiguously associated with the 
true `core' rather than with larger scale `jets')
and black hole mass
or bulge luminosity for a sample dominated by nearby LLAGNs.
It is possible that the range of actual Eddington 
ratios \ledd (which could be responsible for the
scatter in radio luminosities at any given black hole mass) is 
small enough among LLAGNs in our sample to not destroy the correlation between sub-parsec 
radio luminosity and black hole mass or bulge luminosity.
Indeed, the estimated \ledd\ for the nuclei in Fig.~\ref{figvlbamdo} (calculated 
assuming the jet power, Q$_{\rm jet}$, dominates the accretion energy output; see 
Sect.~\ref{secjetpower}) spans a relatively narrow range (see points and dashed 
lines in the left panel of Fig.~\ref{figvlbamdo}).
\citet{ho02}, on the other hand, used a sample with a larger range of \ledd\ and 
measured nuclear radio luminosities at lower resolution (5{\arcsec}; or several hundreds of 
parsecs). Both factors (`classical' Seyferts are apparently more likely to produce radio 
emission on 100-pc scales than LLAGNs; see Sect.~\ref{secjets}) 
could have lead to the absence of a correlation between radio luminosity and black hole mass
in his sample.
Similar factors would also apply to the results of \citet{woourr02}, who used a sample 
which covered a larger range of \ledd\ and which 
included many powerful (and presumably relativistically beamed) radio sources.

\subsection{Radio Luminosity Function}
\label{secrlf}

\begin{figure*}[ht]
\resizebox{\textwidth}{!}{
   \includegraphics{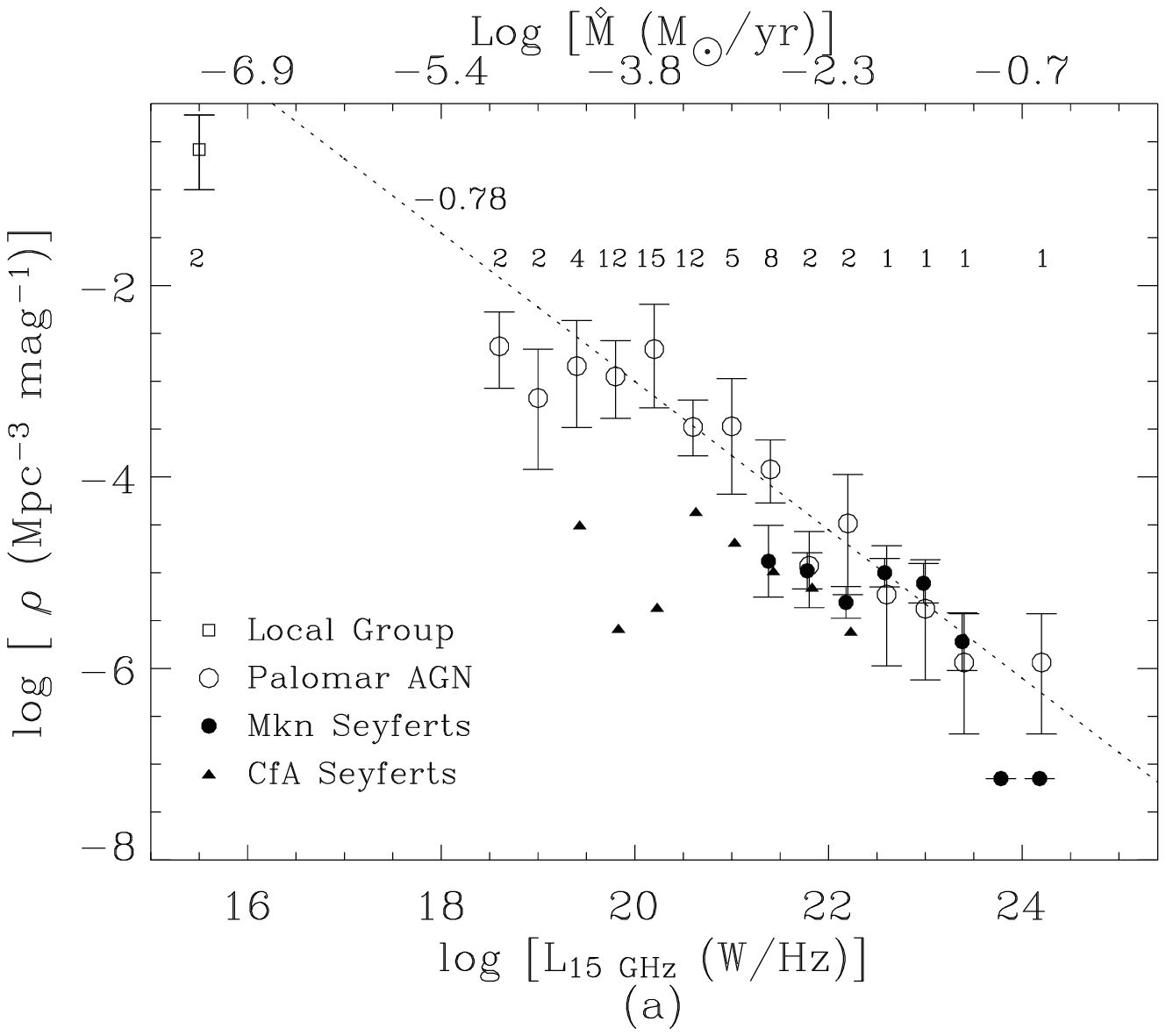}
   \includegraphics{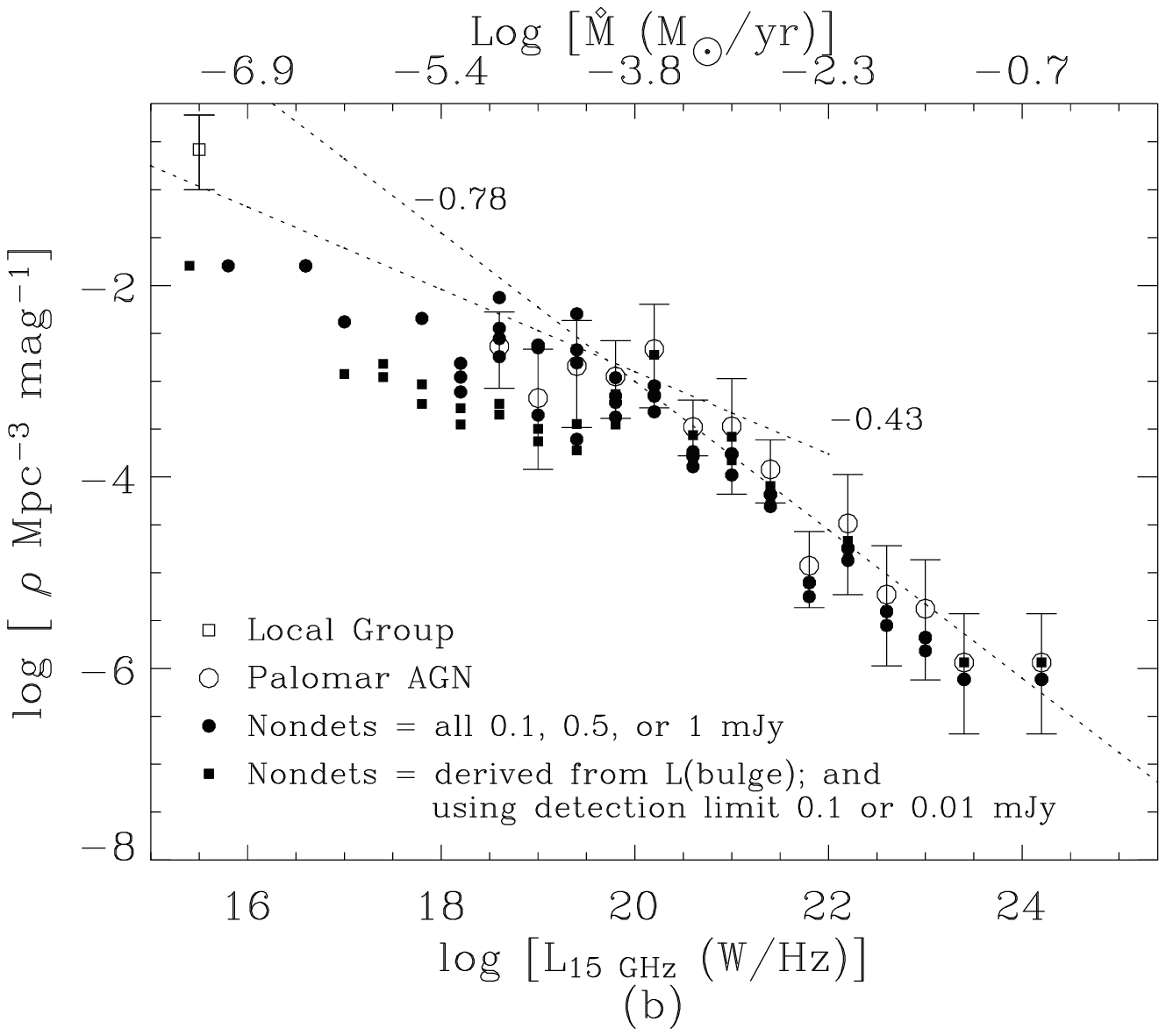}
}
\caption{\textbf{(a)}The 15~GHz radio luminosity function (RLF) of the 150~mas-scale
radio nuclei in the LLAGNs and AGNs of the Palomar sample  
(open circles, with the number of galaxies in each 
bin listed above the symbol). For a rough comparison (see text) we also plot the 
15~GHz RLFs (after converting to our value of H$_0$ and frequency; see text) of
Markarian Seyferts and CfA Seyferts. 
The dotted line is a power-law ($-0.78$) fit to the Palomar nuclear 
RLF (excluding the two lowest radio luminosity points).
Also shown is the estimated 15~GHz nuclear RLF of galaxies in the
local group (open square, with 2 galaxies; see text).
The upper $x$ axes of both panels show the implied logarithm of the mass
accretion rate (in M$_{\odot}$ yr$^{-1}$) within the context 
of a jet model \citep{falbie99}, assuming both a 10\% energy conversion efficiency,
and that the jet kinetic power dominates the accretion energy output (see Sect.~\ref{secjetpower});
\textbf{(b)} the axes and open symbols are the same as in (a). 
 The filled symbols show several simulated nuclear RLFs for the Palomar sample 
 which include nuclei not detected in our 15~GHz survey in the RLF calculation
 (see text for details). 
 The dotted line with slope $-$0.43 shows a possible fit (made by eye)
 to the RLF at the lowest luminosities; the other dotted line is the same as in
 the left panel.
}
\label{figrlf}
\end{figure*}

The nuclear (150~mas-scale) 15~GHz RLF for all 68 radio-detected Palomar 
sample LLAGNs and AGNs (Table~\ref{tabvla}) is plotted in Fig.~\ref{figrlf}a 
as open circles. LLAGNs not detected in the radio have been excluded from the 
RLF calculation.
Only 2 of the radio detections in the Palomar sample (NGC~1275 and NGC~4151) are 
true AGNs as defined by their emission-line properties 
i.e. they have L$_{\rm H\alpha}$ $>>$ 10$^{40}$ erg s$^{-1}$.
The \hii-region nuclei
and absorption-line nuclei, which are excluded from this RLF calculation, 
have much lower radio luminosities than the LLAGNs and AGNs in the sample. 
The RLF has been computed via the bivariate optical-radio luminosity 
function \citep[following the method of][]{meuwil84}, after correcting for
the incompleteness \citep{sanet79} of the RSA catalog (from which the
Palomar sample was drawn). Errors were computed following the method of 
\citet{con89}.
We emphasize that the nuclear 15~GHz RLF presented here traces
only the inner AGN jet or accretion inflow, and does not include the
contribution from $>$150~mas-scale radio jets.

RLFs at 1.4~GHz and 5~GHz for Palomar Seyferts have been presented in \citet{ulvho01a},
and a RLF (using observations at several frequencies and resolutions) for the complete 
Palomar sample has also been discussed in Filho (2003, $PhD$ thesis).
The RLF we present here \citep[first presented in][]{nag03}
is in rough agreement with the above RLFs given the errors.
The advantages of the RLF presented here are threefold.
First, it is based on a larger number (68) of radio detections.
Second, it is derived from uniform radio data: all except 13 radio detections
and 21 radio non-detections have their fluxes or upper limits derived from our 
15~GHz (2~cm) VLA A-configuration observations reduced in a uniform way; these
34 exceptions have fluxes or upper limits derived from data of similar 
resolution and frequency.
Third, the radio data were obtained at high resolution and high frequency: 
both these factors reduce the contamination of star-formation-related
emission to the true AGN radio emission, which is especially important at
these low AGN luminosities.

At the highest luminosities the Palomar RLF is in good agreement with
that of `classical' Seyferts (Fig.~\ref{figrlf}a),  
as previously noted by \citet{ulvho01a}.
We have plotted the RLFs of Markarian Seyferts \citep[1.4~GHz RLF from][]{meuwil84}
and of CfA Seyferts (1.4~GHz RLF calculated by \citet{ulvho01a} using 8~GHz 
data from \citet{kuket95}), after conversion to our values of 
H$_0$ and frequency (assuming that the 1.4~GHz emission is optically-thin
with spectral index $-$0.75). Of course, the `classical' Seyfert RLFs are not 
strictly comparable to ours since our 15~GHz survey detected flat-spectrum 
emission \citep{naget01,naget02b} which may have been invisible to
the 1.4~GHz observations, and conversely, the 15~GHz observations
may not have detected the steep spectrum emission which dominated
the 1.4~GHz observations.
Furthermore, the AGN-related radio structures in the Palomar LLAGNs are either 
sub-arcsec (i.e. the nuclear radio emission is the total AGN-related radio emission) 
or, in a few cases, FR~I-like.  Neither of these can be easily compared or corrected 
to the radio structures seen in most Markarian or CfA Seyferts at lower 
radio frequencies.

At lower luminosities, the sample extends the RLF of powerful AGNs by more than 
three orders of magnitude. A linear (in log-log space) fit to the Palomar nuclear RLF
above 10$^{19}$ Watt Hz$^{-1}$ (i.e. excluding the two lowest luminosity bins; see below) 
yields:
\begin{eqnarray}
{\log}\,\rho &=&\left(12.5 - 0.78\times\,{\rm log}\,\left({\rm L}_{\rm 15\,GHz}\,[{\rm W\,Hz}^{-1}]\right)\right)\nonumber\\ 
  &&[{\rm Mpc}^{-3}\,{\rm mag}^{-1}]
\label{eqnpowlaw}
\end{eqnarray}

As we discuss below, a potential fit to the RLF below 10$^{19}$ Watt Hz$^{-1}$
is (with units as in eqn.~\ref{eqnpowlaw}):
\begin{equation}
{\rm log}\,\rho\, = 5.7\, -\,0.43\,\times\,{\rm log}\,{\rm L}_{\rm 15\,GHz}
\label{eqnbreak}
\end{equation}

There is some indication of a low luminosity turnover in the Palomar RLF
(Fig.~\ref{figrlf}a). Admittedly, this apparent turnover is
partly due to the incompleteness of the radio survey, i.e. biased by
the sub-milli-Jansky population which remains undetected.
Nevertheless there are several reasons to believe the presence of such
a turnover, as detailed below and in Fig.~\ref{figrlf}b.
First, and most convincingly, one runs out of bright galaxies:
an extension of the $-0.78$ power law fit to lower luminosities would require
e.g. an LLAGN like Sgr~A* or M~31* to be present in every Mpc$^{-3}$.
However, there are only about 318 known non-irregular galaxies at 
D$\,<\,$10~Mpc of which only 70 have log~(L/\Lsun)$\,>$ 9 and only 
22 have log~(L/\Lsun)$\,>$ 10 \citep{karet99}.
To better determine the RLF shape at lower luminosities, we have calculated 
an approximate RLF{\footnote{Here we use the highest resolution 
 radio flux, instead of a `matched' linear-resolution flux. The latter flux
 is dominated by extra-nuclear (non-AGN related) emission for these lowest
 luminosity AGNs and is thus not directly comparable to the radio fluxes of the
 other more luminous radio sources used in the RLF computation.}} 
for the nuclei of the local group of galaxies.
The local group has 13 galaxies with M$_{\rm v}\,<\,-14.8$, of which
6 are classified as Irregular. The remaining 7 have been surveyed
at high resolution with the VLA to detection limits between 
30~$\mu$Jy and 1~mJy. 
Our Galaxy \citep[$\sim$1~Jy for Sgr~A*;][]{kriet98} and M~31 
\citep[0.033~mJy][]{craet93} have nuclear radio luminosities 
$\sim10^{15}$ Watt Hz$^{-1}$, and the other five all have upper
limits to their nuclear radio flux:
M~32 \citep[NGC~221, $<\,0.03$~mJy;][]{hoet03b}, 
NGC~185 \citep[$<\,0.12$~mJy;][]{houlv01},
NGC~147 and NGC~205 \citep[$<\,$1~mJy;][]{hecet80},
and NGC~598 ($<\,0.03$~mJy from our reduction of 8.4~GHz A-configuration 
VLA archive data from project AC342). 
We calculate the local group RLF using a simple V/V$_{\rm max}$ 
test for the Galaxy and M~31, assuming a detection limit of 30~$\mu$Jy.
This local group point, plotted as an open square
in Fig.~\ref{figrlf}a and b, is also consistent with
a low-luminosity break in the overall RLF. The error bar on the local group RLF point
is from Poisson statistics; the true error, from the generalization of
our (a-posterior) local group properties, is likely to be larger.

As a further test we simulated the shape of the Palomar RLF at low luminosities
by converting some or all of the LLAGNs not detected in our 15~GHz survey into radio 
detections as follows (a total of 6 simulated RLFs; filled symbols in Fig.~\ref{figrlf}b).
We first recomputed the RLF assuming that all 125 radio non-detected LLAGNs had 
radio nuclei with flux 0.5~mJy and using this value as the assumed detection limit
of the survey.
We then recomputed the RLF for the three cases that the non-detected Seyferts
and LINERs (71 nuclei) all had radio nuclei with flux 1~mJy, 0.5~mJy, or 0.1~mJy 
(and using the corresponding flux as the assumed detection limit of the survey).
The above four simulated RLFs are plotted with filled circles in Fig.~\ref{figrlf}b.
To explore two more possibilities, we set the radio flux of 
individual non-detected LLAGNs to the values expected from the rough proportionality 
between bulge luminosity and 150~mas-scale radio luminosity for nearby galaxies 
\citep[dashed line in Fig.~18, lower panel, of][]{naget02a}.
We then recomputed the RLF for two assumed detection limits (0.1~mJy and 0.01~mJy) and 
ensured that the estimated fluxes fell in the range between 1~mJy (our survey's actual 
detection limit) and the assumed detection limit (if the estimated flux was lower 
than the assumed detection limit then the nucleus continued to be treated as a 
non-detection).
The resulting two simulated RLFs, calculated from a total of 98 and 170 radio 
detections, respectively, are plotted with filled squares in Fig.~\ref{figrlf}b. 
All the simulated RLFs support a low luminosity break in the Palomar RLF.
The actual shape of the low end of the RLF is uncertain and in Fig.~\ref{figrlf}b
and Eqn.~\ref{eqnbreak} we show a potential power law fit which satisfies the 
current data and extrapolations. 

\subsection{Sub-parsec Jets}
\label{secjets}

About 20 of the 44 sources in Table~\ref{tabvlba} have detected sub-parsec scale 
(and/or sometimes larger scale) `jets'. These include
NGC~315  \citep{cotet99b,fomet00,gioet01},
NGC~1167 \citep[][this work]{gioet01},
NGC~1275 \citep{dhaet98,walet00},
NGC~2273 (this work),
NGC~3031 \citep[M~81;][]{bieet00},
NGC~3079 \citep{troet98},
NGC~4151 \citep{munet03,ulvet98},
NGC~4258 \citep{heret97},
NGC~4261 \citep[e.g.][]{jonet01},
NGC~4278 \citep{jonet84,falet00},
NGC~4374 \citep[M~84;][this work]{wroet96},
NGC~4486 \citep[M~87;][]{junbir95},
NGC~4552 \citep[M~89;][]{naget02a},
NGC~4589 and NGC~5353 (this work),
NGC~5354 \citep{filet04},
NGC~5363 (this work),
NGC~5846 \citep{filet04},
NGC~6500 \citep{falet00}, 
and NGC~7626 \citep[this work, also tentatively detected by][]{xuet00}.

The mas-detected radio nuclei fit into four 
categories, with Seyferts and LINERs preferentially belonging to one or the
other category:
(a)~powerful radio galaxies or low power radio galaxies --
 NGC~315, 
 NGC~1275, 
 NGC~4261, 
 NGC~4374, 
 NGC~4486 and 
 NGC~7626 
 -- which have an elliptical host, a LINER nuclear spectrum, and collimated 
 sub-parsec to kpc jets;
(b)~nuclei which do not have detected parsec-scale jets, but have larger 
 (100~pc- or kpc-scale) jets. These are preferentially Seyfert-like nuclei 
 (6 nuclei) though NGC~5363 (LINER) and NGC~5846 (transition nucleus) fall in 
 this class;
(c)~nuclei with detected sub-parsec jets but weak or no known larger scale jets  
    - preferentially in LINERs or transition nuclei. These nuclei --
 NGC~4278,
 NGC~4552,
 NGC~4589,
 NGC~5353,
 NGC~5354, and
 NGC~6500 --
 typically show curved or highly bent jets \citep[][this work]{naget02a,filet04}
 and proper motion studies (Nagar et al., in prep) suggest that these jets are
 frustrated in the inner few parsecs;
(d)~the remaining $\sim$23 nuclei in Table~\ref{tabvlba} do not show extended mas-scale 
 emission but require more detailed study of their 100~pc to kpc scale radio emission.
 The disappearance of the larger-scale jet in category (c) above, and the
 morphology of their parsec-scale jets (highly curved or bent) suggests that 
 the jet does not propagate beyond the inner few parsecs, either due to being 
 uncollimated, or because of interaction with the ambient medium. If this is 
 the case, the energy deposited into the inner few parsecs by the jet (next 
 section) is significant, and could potentially be responsible for lowering the larger 
 scale (i.e. outside the accretion disk) gas inflow and thus ultimately the accretion 
 rate.

\subsection{Jet Power versus Radiated Luminosity}
\label{secjetpower}

\begin{figure}[ht]
\resizebox{\columnwidth}{!}{
  \includegraphics{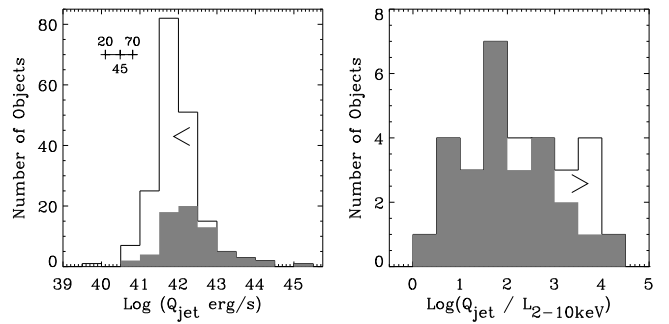}
}
\caption{\textbf{Left:} the implied `minimum jet power', 
(Q$_{\rm jet}$) of the radio-detected 
(grey shaded area) and radio non-detected (white area) LLAGNs and AGNs, 
calculated from the peak VLA 15~GHz flux using Eqn. 20 
of \citet{falbie99} and assuming a jet inclination of 45{\degr} to the line of sight. 
The inset illustrates the range of calculated minimum jet kinetic powers for three
assumed inclinations: 20{\degr}, 45{\degr}, and 70{\degr}.
\textbf{Right:} log of the ratio of minimum jet power (assuming 45{\degr} jet inclination)
to X-ray luminosity (in the 2--10~keV band) for
radio detected LLAGNs and AGNs. The grey and white histograms represent LLAGNs/AGNs with 
detected, and upper limits to, the hard X-ray emission, respectively.
}
\label{figjettoxray}
\end{figure}

\begin{figure}[ht]
\resizebox{\columnwidth}{!}{
  \includegraphics{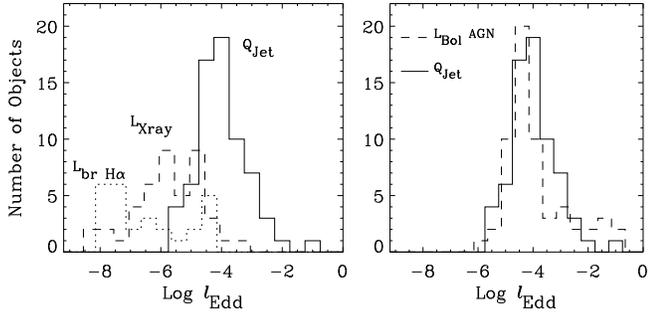}
}
\caption{A comparison of the kinetic and radiated accretion power outputs as a fraction of
L$_{\rm Eddington}$ in four energetically important wavebands.
In both panels the histograms are offset by 0.1 in $x$ for clarity.
\textbf{Left:} histograms of the
broad H$\alpha$ luminosity (dotted line), 
hard X-ray luminosity (2--10~keV; dashed line),
and minimum jet power (Q$_{\rm jet}$; solid line)
for all broad line nuclei, all hard X-ray detected nuclei, and all radio detected nuclei,
among the LLAGNs and AGNs of the Palomar Sample, respectively. 
\textbf{Right:} histograms of the minimum jet power (Q$_{\rm jet}$; solid line) and
radiated bolometric luminosity (derived from the \oiii\ luminosity, see text; dashed line)
for the radio detected nuclei in the LLAGNs and AGNs of the Palomar Sample. 
}
\label{figmultiband}
\end{figure}

\begin{figure*}[ht]
\sidecaption
\resizebox{5in}{!}{
  \includegraphics{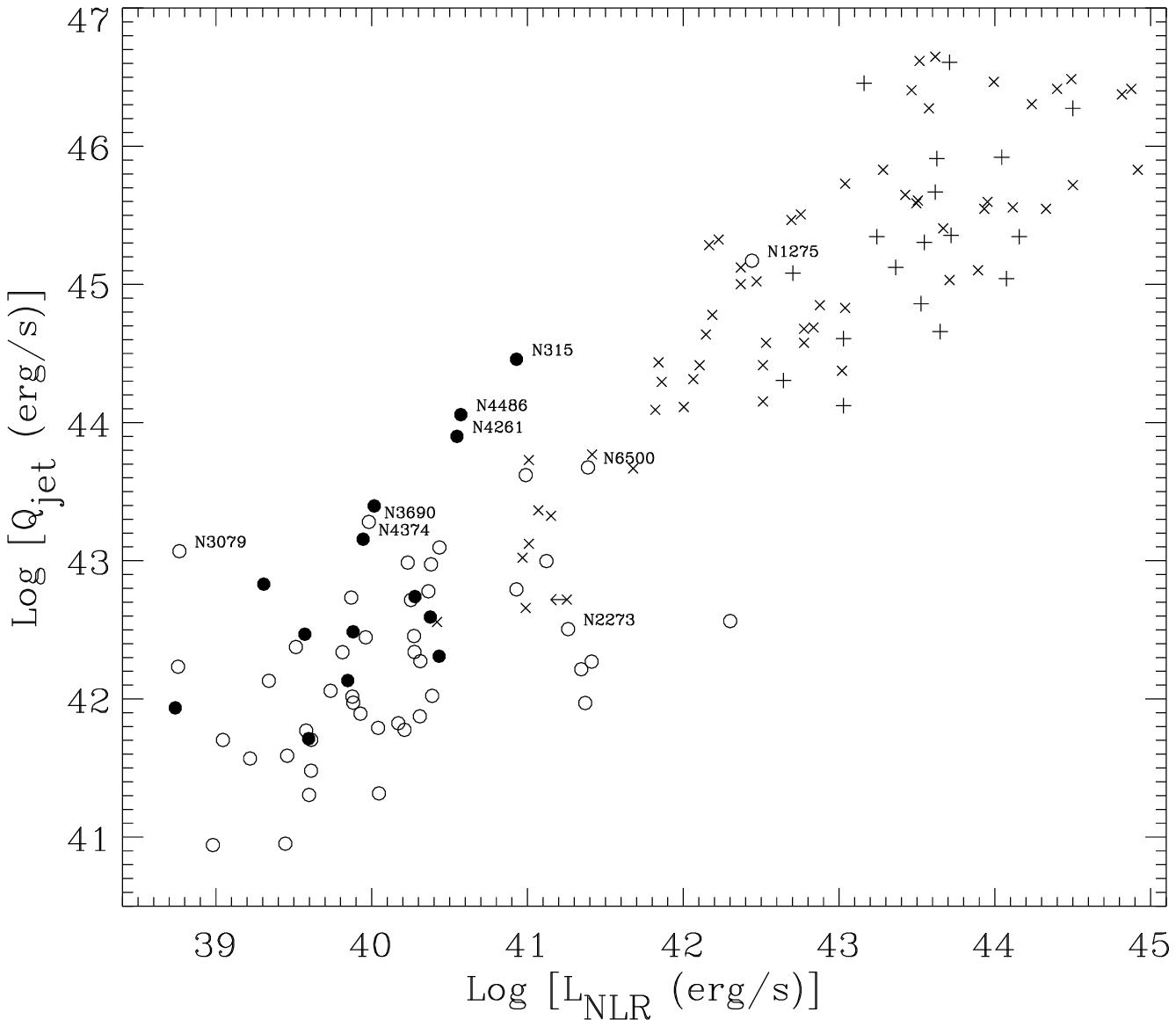}
}
\caption{A plot of Jet Power (we use the `minimum jet power', Q$_{\rm jet}$, derived 
  from the \textit{nuclear} radio luminosity; see text) versus the Narrow Line Region (NLR) 
  luminosity (derived from the \oiii\ luminosity; see text)
  for radio-detected nuclei in the Palomar Sample (circles). Filled circles are used for 
  elliptical galaxies in the Palomar sample. These Palomar Sample LLAGNs and AGNs nicely fill in 
  the low luminosity end of similar relationships for FR~I and FR~II radio galaxies
\citep[slanted crosses;][with jet kinetic energy estimated from their kpc-scale radio lobes]{rawsau91},
  and other radio galaxies 
\citep[crosses;][with jet kinetic energy estimated from their parsec-scale radio jets]{celfab93}.
The data from other papers have been corrected to the value of H$_0$ used in this paper.
}
\label{figcelotti}
\end{figure*}

\subsubsection{Jet Power Domination in Palomar LLAGNs}

With an estimated black hole mass and an emitted luminosity, one can estimate the 
Eddington ratio, i.e.  \ledd\,= L$_{\rm Emitted}$/L$_{\rm Eddington}$.
Previous calculations of \ledd\ for LLAGNs have considered only the radiated component 
of L$_{\rm Emitted}$. 
Since LLAGNs lack a `big blue bump', the X-ray emission has been thought to dominate 
the bolometric luminosity \citep{ho99}. 
With most LLAGNs having hard X-ray luminosities of only $\sim10^{40}$ erg s$^{-1}$ 
or lower, the accretion is inferred to be highly sub-Eddington 
\citep{hoet01,terwil03,filet04}.

If, as justified above, the compact radio nuclei and sub-parsec jets represent
emission from the base of a relativistic jet launched close to the black hole,
then the kinetic energy in the jet can be quite high.
Equation 20 of \citet{falbie99} - assuming an average inclination
of 45{$\degr$} - can be used to obtain `minimum jet powers' (Q$_{\rm jet}$)
of $10^{40}-10^{45}$ $\ergsec$ (Fig.~\ref{figjettoxray}, left panel) 
from the 15~GHz peak VLA flux of radio detected LLAGNs. 
For LLAGNs with both hard X-ray and radio luminosity available, this jet power 
greatly exceeds the radiated X-ray luminosity (Fig.~\ref{figjettoxray}, right panel). 
Since the bolometric luminosity (L$_{\rm Bol}$)
in electromagnetic radiation is estimated to be only 
$\sim3-15\,\times$ L$_{0.5-10~\rm keV}$ for LLAGNs \citep{ho99}, this suggests 
that the accretion power output is dominated by the jet power. 
This domination of jet power over X-ray emission is analogous to the situations for
$\sim$10-15 \msun\ black holes in Galactic X-ray binary systems \citep{fenet03}
and for powerful radio galaxies \citep[e.g.][]{celfab93,oweet00}.

To expand on this issue, we compare the estimated minimum jet power to the observed 
hard X-ray luminosity and emission-line luminosities for the Palomar sample LLAGNs.
Clearly (Fig.~\ref{figmultiband}, left) the minimum jet power is significantly larger
than the measured hard X-ray luminosity (as noted above) and the luminosity in broad \ha.
On the other hand, if the radiated bolometric luminosity is estimated from the \oiii\
luminosity -  from the empirical result that the spectral energy
density of type~1 AGNs typically shows
L$_{\rm Bol}\,= 3500 \times {\rm L}_{\rm [OIII]}$ 
\citep[][see also next Sect.]{hecet04} -
then LLAGNs show similar distributions of minimum jet power and radiated bolometric 
luminosity (Fig.~\ref{figmultiband}, right).

\subsubsection{A Common Jet-Power Scaling for LLAGNs and Powerful AGNs}

The radio detected LLAGNs in the Palomar sample (circles; Fig.~\ref{figcelotti})
fall nicely at the low luminosity end of the correlation between jet kinetic 
power and total emission line luminosity from the narrow line region (NLR) 
for more powerful FR~I and FR~II galaxies. This figure includes FR~I and FR~II 
radio galaxies \citep[slanted crosses; in these the jet kinetic power is estimated from 
lobe-feeding energy arguments;][]{rawsau91} and other powerful radio sources 
(including BL Lacs, quasars, and radio galaxies) with parsec-scale jets 
\citep[crosses; in these the bulk kinetic energy in the jet was estimated using 
a self-Compton synchrotron model applied to the parsec-scale jet;][]{celfab93}.
We estimated the NLR luminosity for the Palomar LLAGNs and
AGNs following \citet{celfab93}: \newline 
L$_{\rm NLR}\,=\,3\,\times\,(3\,$L$_{\rm [OII]\lambda3727}\,+\,1.5\,$L$_{\rm [OIII]\lambda5007}$) \newline
with L$_{\rm [OII]\lambda3727}$ estimated as $0.25 \times$ L$_{\rm [OIII]\lambda5007}$ for Seyferts
and $3 \times$ L$_{\rm [OIII]\lambda5007}$ for LINERs and transition nuclei. The Palomar \oiii\ 
luminosities were measured in a 2{\arcsec}$\times$4{\arcsec} nuclear aperture \citep{hoet97a}; 
these values closely approximate the total NLR luminosity in most nuclei except a handful, 
e.g.  NGC~4151, which have more extended NLRs.

\subsubsection{LLAGN Jet-Power and Global Energetics}
\label{secglobalenergetics}
\begin{figure}[ht]
\resizebox{7cm}{!}{
  \includegraphics{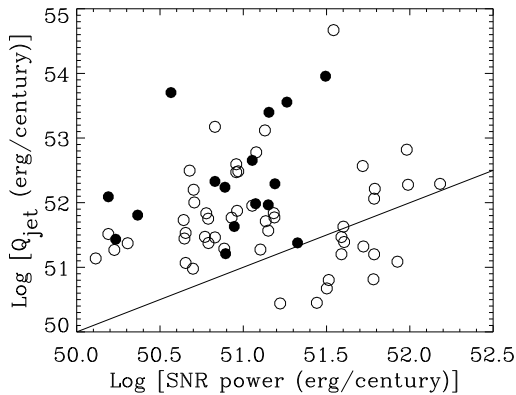}
}
\caption{A comparison of the `minimum jet power' (Q$_{\rm jet}$) and the kinetic energy
injected into the ISM by supernovae type I and II for the radio detected Palomar 
LLAGNs and AGNs. Filled circles are used for elliptical galaxies. The solid line shows
the line of equality.
}
\label{figsnr}
\end{figure}

The energetics of the jet are also important in the context of so called cooling flows 
and in regulating the feedback between galaxy growth and black hole growth. For example,
in many clusters the central cD galaxy has an FR~I radio morphology, with the radio jet 
playing an important role in the above issues and in the global energetics of the cluster 
gas \citep{oweet00,bin04b,ostcio04}.
A comparison of the jet kinetic power with the power
injected into the ISM by supernovae
types I and II in the host galaxies 
is shown in Fig.~\ref{figsnr} for the radio detected 
Palomar LLAGNs/AGNs.
The supernova rates (as a function of galaxy morphological type) are taken from 
the ``Case B'' values in Table 6 of \citet{vanmcc94},
e.g. $0.25\,$SNu for E and S0 galaxies (all from SN type Ia), 
where 1 SNu = 1 SN per century per 10$^{10}$(L$_{\rm B}$/L$_{\sun}$). The total SN rate is
slightly higher in later type galaxies due to the contribution of SN type II.
Values of L$_{\rm B}$/L$_{\sun}$ were taken from \citet{hoet97a} and each supernova is assumed 
to inject 10$^{51}\,$erg of kinetic energy into the ISM \citep[e.g.][]{bintre87,pelcio98}.
The jet power is clearly the major player in the nuclear energetics not only because it
exceeds the total SN kinetic power in almost all cases, but also because its nuclear origin 
allows a closer `feedback' to the accretion inflow. 
A significant fraction of the jet energy is potentially deposited in the central parsecs,
especially in LLAGNs which show pc-scale (usually bent) jets but no larger scale jets 
(Sect.~\ref{secjets}); such deposition of jet power  
can considerably slow down or balance any cooling flow or
other inflow in the inner parsecs and thus ultimately help `starve' the
accretion disk
\citep[e.g.][]{dimet01}.
Additionally, LLAGNs with kpc-scale jets inject significant energy into the inter-galactic
medium (IGM), and work against any cooling flow \citep[see e.g.][]{pelcio98}.
The most recent of such `feedback' analyses \citep{bin04b}
does take into account the jet
power, though for the more powerful FR~I type jets in cD galaxies. Our results 
show that their models can be applied, at least qualitatively, to LLAGNs.

In summary, Eddington ratios,  
\ledd, calculated from hard X-ray luminosities 
heavily underestimate the true accretion power output in LLAGNs.
This finding is in line with that for more powerful radio galaxies, where the
jet kinetic power is two to five orders of magnitude larger than the radiated
radio luminosity, and often significantly larger than the total radiated bolometric 
luminosity \citep{celfab93,oweet00}.
Using L$_{\rm Bol}$ = 3500 L$_{\rm [OIII]}$ or L$_{\rm Bol}$ = Q$_{\rm jet}$ 
yields similar distributions of L$_{\rm Bol}$.
This, and the scaling between L$_{\rm NLR}$ and Q$_{\rm jet}$,
argues for a common central engine in all AGNs from LLAGNs to powerful 
FR~IIs, but with the caveat that the \oiii\ luminosity in LLAGNs is potentially 
contaminated by non accretion related processes. Finally, the jet is potentially
a significant (maybe even dominant) source of heating in the galaxy. If the jet
is disrupted in the inner parsecs, then the jet power could play a role in slowing
any cooling flow or other accretion inflow on parsec-scales, thus starving the 
accretion disk.

\subsection{The Eddington Ratio}
\label{seceddington}

In the previous section we showed that the accretion energy output in LLAGNs with 
radio nuclei is dominated by the jet power, and is of the order of 
\ledd\,= 10$^{-6}$ to 10$^{-2}$. 
Here we look at the dependence of this jet-power-derived \ledd\ on other quantities.

\citet{hecet04} have investigated black hole and galaxy growth using 23,000 type~2
AGNs from the Sloan Digital Sky Survey (SDSS). They use $\sigma_c$ to estimate
black hole mass, and L$_{\rm Bol}$ = 3500 L$_{\rm [OIII]}$.
They find that most present-day accretion occurs onto black holes with masses 
$<\,3\,\times 10^7$ \Msun, and that most black hole growth takes place in systems with 
accretion rate less than one fifth of the Eddington rate.
It is interesting to apply their analysis to the Palomar LLAGNs since these objects
have emission-line luminosities typically ten to a hundred times fainter than the 
SDSS AGNs. 
The Palomar LLAGNs show a different behavior to the SDSS AGNs:
the numerous Palomar LLAGNs with low black hole masses are accreting
a similar 
mass per year as the fewer Palomar LLAGNs and Palomar AGNs at higher radio luminosity.
This is true whether the accretion rate is calculated from the \oiii\ 
luminosity, as in \citet{hecet04}, or from Q$_{\rm jet}$
(e.g. Fig.~\ref{figrlf}, upper $x$-axis). 

\begin{figure*}[ht]
\resizebox{\textwidth}{!}{
\includegraphics{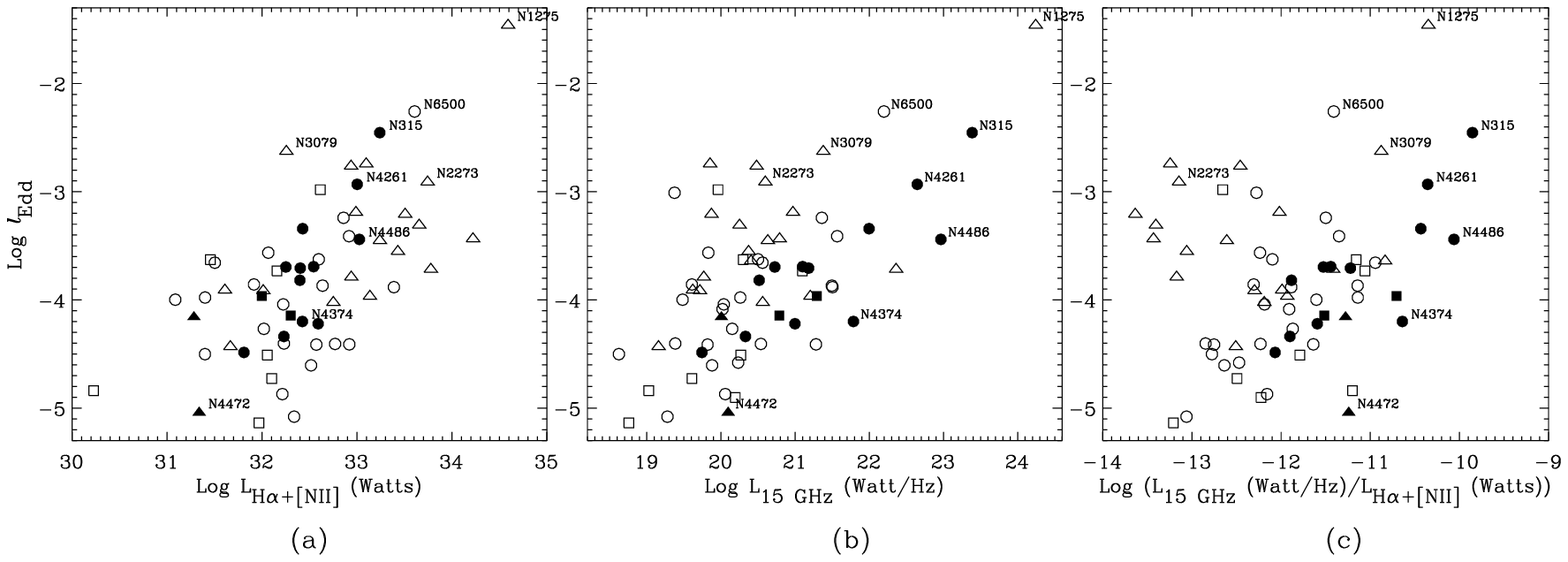}}
\caption{
 Plots of the `minimum jet power' as a fraction of Eddington luminosity 
 (equivalent to the Eddington ratio, \ledd, if the jet power dominates the emitted 
 accretion energy) versus  
   \textbf{(a)}~the nuclear \fullhanii\ luminosity, 
   \textbf{(b)}~the 15~GHz nuclear (150~mas resolution) radio luminosity, and
   \textbf{(c)}~ratio of the above two quantities, 
 for all radio-detected LLAGNs and AGNs in the Palomar sample.
 Seyferts are plotted as triangles, LINERs as circles, and transition nuclei
 as squares. Filled symbols are used for elliptical galaxies.}
\label{figeddington}
\end{figure*}

The Palomar sample ellipticals (filled symbols in Fig.~\ref{figeddington})
show a strong correlation between Eddington
ratio,  \ledd\ (calculated assuming the jet power dominates the 
accretion energy output)
and all of emission-line luminosity, radio luminosity,
and the ratio of radio luminosity to emission line luminosity. 
Among the non-elliptical nuclei (open symbols in Fig.~\ref{figeddington}),
Seyfert nuclei (triangles) display higher Eddington ratios
than LINERs (circles), even 
though the distribution of radio luminosities (from which Q$_{\rm jet}$ is calculated) is 
similar for the two classes.
This is most clearly noticeable in Fig.~\ref{figeddington}c.

\subsection{Discovery of LLAGNs: A Comparison of Radio, Optical, and X-ray
Methods}
\label{secident}
The current study and sample allows a comparison of the relative success 
of deep optical spectroscopy, high resolution radio imaging, and (to some
extent) hard X-ray imaging, in identifying low-luminosity accreting black holes.  
Such a comparison is specially relevant in view of the several hundred thousand 
AGNs and LLAGNs being identified by current large surveys out to
$z \sim$6.
We consider three factors here: 
efficiency, reliability, and completeness. In this subsection, we shall
use the term ``AGN'' to mean ``AGN + LLAGN'', namely objects powered by 
accretion onto a supermassive black hole. 

The optical spectroscopy (of $\sim$486 nuclei) was obtained at the 5~meter Hale 
telescope with typical exposures of 30~min to 1~hr per nucleus.
Has this optical spectroscopy missed any (radio- or X-ray identified) 
AGNs 
in the Palomar sample? The spectroscopic survey found emission-lines in all 
except 53 nuclei \citep{hoet03a}. 
Of these 53 absorption line nuclei we found only two with radio nuclei which would 
identify them as definite AGNs, and a further three to five with radio nuclei possibly
related to an AGN (see the appendix).  
We could not find reliable hard
X-ray identifications of AGNs, for lack of data, in 
any of the absorption line nuclei. The \hii\ nuclei in the Palomar sample also
show no indication of an AGN in the radio \citep{ulvho01b} or X-ray.
Thus, the optical spectroscopic survey has missed only two 
definite (and perhaps three or five probable) AGNs which would have been picked 
up by a radio survey with detection limit $\sim$1~mJy.
The reliability of AGN identification from optical spectroscopy alone is hard
to quantify.
At these low luminosities, emission lines could be
powered by sources other than an AGN, e.g. a nuclear starburst \citep{maoet98}. 

\begin{figure}
\begin{center}
\resizebox{2.7in}{!}{
\includegraphics{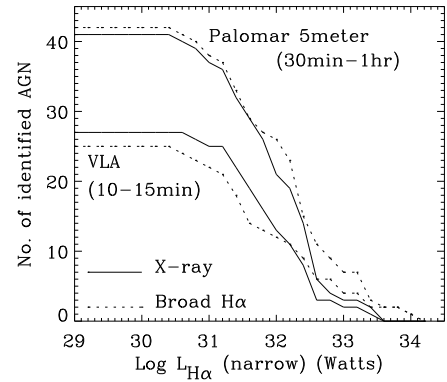}}
\end{center}
\caption{The cumulative number of `definite' 
(solid lines for nuclei with a hard X-ray nuclear source and 
dashed lines for type 1 nuclei, i.e.  with broad H$\alpha$ emission) 
AGNs identified by optical spectroscopy (upper two lines; 
$\sim$30~min to 1~hr on the Palomar 5~meter)
and radio imaging (lower two lines; $\sim$10--15~min at the VLA), 
as a function of the luminosity of the narrow H$\alpha$ line. 
}
\label{figoptvsrad}
\end{figure}

The radio imaging (of $\sim$200 nuclei) was done with the VLA with typical integration 
times of 10 to 15~min per nucleus. Follow-up VLBA imaging (1~hr per
nucleus) showed that the VLA-only imaging could be used for reliable identification
of the radio nuclei as AGN-related (Sect.~\ref{secvlbares}).
The high brightness temperature radio nuclei and parsec-scale jets,
found through VLBA observations, are the 
most reliable indicators of AGNs in these nuclei as a class.
Our radio imaging has identified fewer AGN candidates 
than the optical  spectroscopy, 
though it is likely that deeper radio imaging will uncover significantly
more AGNs \citep{naget02a}. 
It is difficult to ascertain whether the radio survey has missed any 
(definite) optically identified AGNs, since the detection of weak emission lines does not 
guarantee the presence of a low-luminosity accretion-powered nucleus.
Ideally, one requires hard X-ray confirmation of the presence of an AGN: this issue
will be more fully discussed in Terashima et al., (in prep.).
The cumulative number of `definite' AGNs identified by the optical spectroscopic
and the radio imaging methods, as a function of the luminosity 
in the narrow 
H$\alpha$ line, is shown in Fig.\ref{figoptvsrad}. Here we use the presence of either a 
nuclear hard X-ray source (solid lines; Fig.~\ref{figoptvsrad}) or broad H$\alpha$ emission
(dashed lines; Fig.~\ref{figoptvsrad}) as the signpost of a `definite' AGN  - 
i.e. an object powered by accretion onto a supermassive black hole.
Our radio survey has detected 26 of 39 (66\%) of the type~1 nuclei in the sample. Of the 14 type~1 
nuclei not detected in our radio survey, all 5 Seyferts were found to have radio nuclei 
in the deeper radio survey of \citet{houlv01}, while the nature of the others is unknown.

In summary, optical spectroscopy of the Palomar sample 
has found almost all radio or X-ray identified AGNs 
to have emission-lines: as discussed in the Appendix, very few of the
nuclei with only absorption lines have radio emission likely powered by an
accreting black hole. A caveat here is that the absorption line nuclei have
not been surveyed in precisely the same way as we have surveyed the AGNs and
LLAGNs.
The optical spectroscopic method also finds
many nuclei with emission lines powered by hot stars (\hii\ nuclei)
and these emission lines
can hide weaker emission lines from an AGN.
High resolution, high frequency radio imaging (the present survey)
has detected a smaller
fraction of AGN candidates 
in the Palomar sample than the optical spectroscopic method,
but we have argued that these radio sources are not
related to stellar processes. 
Thus, the presence of a
compact flat-spectrum high brightness-temperature radio core is a more 
reliable indicator of an accreting black hole than the presence of optical
emission lines, 
at least at these low emission line and radio luminosities. 
A hard X-ray nucleus is also an ideal signpost of an AGN. However, high
resolution (i.e. $Chandra$) is required to minimize confusion with X-ray binaries
or Ultraluminous X-ray sources (ULXs), which have similar luminosities to the AGN in 
LLAGNs. Further, a large X-ray survey is very expensive in terms of telescope time.
Instead, scientific results can be efficiently attained by hard X-ray observations of 
subsamples selected to have a compact radio core \citep[][Terashima et al., in prep]{terwil03}.

\section{Discussion}
\label{secdiscussion}

The VLBI results (Sect.~\ref{secvlbares} and Table~\ref{tabvlba}) 
confirm that almost all (38 of 39, or 97\%) LLAGNs and AGNs in the Palomar sample with 
S$^{\rm VLA}_{\rm 15\,GHz} >$ 2.7~mJy
have detected mas-scale or sub mas-scale radio nuclei with brightness-temperature $\gtrsim\,10^7\,$K.  
The only exception is NGC~2655: this nucleus has a steep spectrum at 
arcsec resolution \citep{naget00} and was not detected by us with the VLBA. 
Deeper VLBA/I maps show mas-scale radio nuclei in 
five Palomar LLAGNs with S$^{\rm VLA}_{\rm 15\,GHz} \leq$ 2.7~mJy 
(Table~\ref{tabvlba}).
It is notable that the LLAGNs and AGNs with known ultracompact radio nuclei are 
divided between Seyferts and LINERs in proportion to their relative numbers in 
the Palomar sample (14 and 23, respectively; see Table~\ref{tabvlba}).
Thus the probability of detecting ultracompact radio nuclei in LLAGNs with
Seyfert and LINER spectra is similar.
Nuclear starbursts have a maximum brightness-temperature of 
$\sim\,10^{4-5}\,$K \citep{conet91} while the most luminous known radio 
supernova remnants \citep[e.g.][]{colet01} would have brightness 
temperatures $\leq\,10^{7}$~K even if they were $\leq$~1~pc in extent.
As argued in \citet{falet00}, if the nuclear radio emission is attributed to
thermal processes, the predicted soft X-ray luminosities of LLAGNs would be 
at least two orders of magnitude higher than observed by \asca \citep{teret00}
and \chandra \citep{hoet01,terwil03,filet04}.
Also, as pointed out by \citet{ulvho01a}, single SNRs \citep{colet01} 
or a collection of SNRs \citep{nefulv00} would have radio spectral indices 
(defined by $S$ $\propto$ $\nu^{\alpha}$)
$\alpha\,\sim\,-$0.7 to $-$0.4 rather than the values $\alpha\,\sim$ $-$0.2 
to 0.2 seen in the VLBA-detected LLAGNs 
\citep[][Nagar et al., in prep]{naget01,naget02b,andet04}.
Furthermore, significant flux variability is observed \citep{naget02a}.
Thus, the only currently accepted paradigm which may account for the sub-parsec
radio nuclei is accretion onto a supermassive black hole. In this case,
the mas-scale radio emission is likely to be either emission 
from the accretion inflow \citep{naret00} or synchrotron emission from 
the base of the radio jet launched by the accreting supermassive black hole 
\citep{falbie99,zen97}. 
The latter model is supported by the presence of 
sub-parsec size jets in many of the nuclei, and the radio spectral shape
\citep{naget01,naget02a,naget02b,andet04}.

The radio results imply that a large fraction (perhaps all) of LLAGNs 
have accreting massive black holes. 
If we consider only the detections of mas-scale radio sources, 
then at least 25\%~$\pm$~5\% of
LINERs and low-luminosity Seyferts have accreting black holes.
VLA-detected compact radio nuclei with flux $<\,$2.7~mJy were not investigated 
with the VLBA; in other respects these nuclei are similar to those with 
detected mas scale structure. Thus it is likely that \textit{all} LLAGNs with 
VLA-detected compact radio nuclei (42\%~$\pm$~7\% of LINERs and low-luminosity 
Seyferts) have accreting black holes.
The scalings between radio luminosity, emission-line luminosity, and galaxy
luminosity \citep{ulvho01a,naget02a,filet04} provide evidence that the radio 
non-detections are simply lower luminosity versions of the radio detections.
In fact we find no reason to disbelieve that \textit{all} 
LLAGNs have an accreting black hole.

Interestingly, ultracompact radio nuclei (Table~\ref{tabvlba}) are found almost exclusively 
in massive (M$_{\rm B}$(total) $\leq\,-20$) ellipticals  and in type~1 LLAGNs, or both. 
For massive ellipticals, the high bulge luminosity and black hole mass appear
to be key factors related to the production of a radio nucleus, in light of the 
scalings seen between radio luminosity and these parameters
(see Fig. 4 of present paper and Nagar et al. 2002a).
Among non-ellipticals, the preferential detection of type~1 LLAGNs may result from 
the limited sensitivity of optical and radio observations, which detect broad \ha\ 
and radio nuclei in only the more luminous LLAGNs.
For example, it may be that type 1 LLAGNs are in an outburst phase
during which they temporarily host both broad \ha\ emission and 
a compact radio nucleus. 
Type~2 LLAGNs, on the other hand, may harbor quiescent AGNs which
do not generate sufficient ionizing photons to power the optical emission lines
\citep[e.g.][]{teret00,filet04}.
Instead, their emission lines could be powered by star formation related
processes \citep{maoet98}.
As another alternative, one can invoke the unified scheme \citep{ant93} and 
posit that all LLAGNs have accreting black holes and either
(a)~the radio emission in type~1 LLAGNs is beamed (weakly relativistic jets 
   [$\gamma\,\sim\,2$] can give boost factors of up to $\sim\,$5) and/or
(b)~the 15~GHz radio emission in type 2 LLAGNs is free-free absorbed by a
    `torus'-like structure i.e. $\tau_{\rm 15\,GHz}\,\geq$~1.

The radio and emission line properties of LLAGNs in elliptical galaxies are
consistent with them being scaled-down FR~Is 
(Sect.~\ref{secemi} and Sect.~\ref{secjetpower}), 
confirming earlier such suggestions with smaller samples 
\citep[][Chiaberge et al., in prep.]{naget02a,veret02}.
Additionally, in the context of jet models, the same scaling relationship 
between jet kinetic power and radiated NLR luminosity is followed by
parsec-scale jets in LLAGNs as kpc-scale jets in powerful FR~I and FR~II
radio galaxies (Fig.~\ref{figcelotti} and Sect.~\ref{secjetpower}).

The nuclear environments of low-luminosity Palomar Seyferts are richer in gas than those of 
Palomar LINERs \citep{hoet03a}, as inferred from higher electron densities (n$_{\rm e}$)
and higher internal extinction in the former class. 
We have found that among non-elliptical hosts, LINER nuclei have lower 
Eddington ratios
than Seyfert nuclei (Fig.~\ref{figeddington} and Sect.~\ref{seceddington}). 
Also, we find evidence for a higher incidence of parsec-scale radio jets in LINERs than 
Seyferts (Sect.~\ref{secjets});  at least 
some low luminosity 
Seyferts do show larger (100~pc scale) jets. It is tempting to speculate,
in analogy to Galactic black hole candidates \citep{fenbel04}, that LINERs with
radio nuclei are in a `low/hard' state (low Eddington ratio, lack of inner accretion disk, 
more efficient at launching collimated jets) while low-luminosity Seyferts are in a `high'
state (higher Eddington ratios, less efficient at launching collimated jets).

Luminous Seyferts, the Palomar sample, and the local group of galaxies together allow 
an estimate of the nuclear radio luminosity function over the radio luminosity range 
10$^{15}-10^{24}$ Watt~Hz$^{-1}$, more than five orders of magnitude larger than
previous AGN samples. At the lowest 
luminosities there is tentative evidence for a turnover in the RLF. One must  
therefore reach as far down as the LLAGN regime (but not necessarily lower) 
to completely study the demographics of nuclear accretion.
This point is especially important since larger surveys, e.g. SDSS, probe
accretion in AGNs with \ledd\ one or two orders of magnitude larger than
that in the Palomar sample LLAGNs.

When only the radiated luminosity is considered, LLAGNs have very low inferred 
Eddington ratios. This requires either  a very low mass accretion 
rate or a radiatively inefficient accretion mechanism or both,  and was among 
the original motivations for invoking RIAFs in LLAGNs 
\citep[but see][for an argument against the existence of RIAFs]{bin04a}.
Including the jet power (Sect.~\ref{secjetpower}) in the accretion output 
weakens  the motivation for a RIAF.
First, including the jet kinetic power significantly increases the total (radiated plus 
kinetic) L$_{\rm Emitted}$ and thus the Eddington ratio.
Second, the energy deposited by the jet into the nuclear regions can 
potentially  heat the gas in the inner parsecs and thus decrease  gas supply to 
the accretion disk (Sect.~\ref{secglobalenergetics}). 
Together these two factors weaken the previous preference for RIAFs over
a matter-starved accretion disk plus jet system.  
The absence in LLAGNs of the `big blue bump' and Fe~K$\alpha$ lines
are thus the main remaining motivations for preferring an optically-thin, geometrically 
thick and advection dominated inner accretion structure over the standard 
optically-thick, geometrically thin accretion disk.

Jet models indicate that the dominant form of power output in LLAGNs
is the kinetic power of the jet (Sect.~\ref{secjetpower}). The Eddington
ratios found are \ledd\ $\sim$ 10$^{-2}$ to 10$^{-6}$ 
(Figs.~\ref{figvlbamdo},
\ref{figmultiband}, \ref{figeddington}).
In terms of a mass accretion rate (assuming a 10\% conversion efficiency) this
translates to $\dot{\rm M} = 10^{-1}$ to 10$^{-5}$ \msun\ yr$^{-1}$
(upper $x$-axis of Fig.~\ref{figrlf}). The summed mass accretion rate for all
Palomar LLAGNs and AGNs in Fig.~\ref{figrlf} is 0.4 $\,$\msun\ yr$^{-1}$.
Now, the idealised  jet model of equation (20) of  \citet{falbie99} 
predicts 
log Q$_{\rm jet}\,\propto 0.79\,\times$ log~L$_{\nu}$  (where L$_{\nu}$
is the observed luminosity at radio frequency $\nu$) for a jet inclination of 45{\degr}, 
while the Palomar RLF has the form 
log $\rho\,\propto -0.78 \times$ log~L$_{\rm 15\,GHz}$.
While there are large uncertainties in applying the jet model, it is worth 
remarking
that these scalings imply Q$_{\rm jet}$ $\propto$ $\rho^{-1}$. 
This implies (since Q$_{\rm jet}$ $\propto$ $\dot{\rm M}$ under the assumption
that the jet power dominates the accretion power output) that
the more numerous nuclei with lower radio luminosity are together accreting the
same mass per
year as the fewer nuclei at higher radio luminosity, 
at least down to the probable RLF turnover at log L$_{\rm 15\,GHz}\simeq$ 19
Watt Hz$^{-1}$.
Finally, with individual jet powers of $\sim\,10^{40}-10^{45}$ $\ergsec$,
LLAGN jets provide a significant source of energy 
(Sect.~\ref{secjetpower}) into the galactic ISM and also perhaps the IGM. 
As discussed
in \citet{dimet01} a fraction of this jet power
deposited within the central $\sim$0.1~kpc
would be sufficient to significantly lower the accretion rate at 
least in the case of spherical
accretion. 

\section{Conclusions}
\label{secconclusion}

We have presented the results of our VLA plus VLBA radio survey of 162 LLAGNs
and AGNs from the Palomar sample of nearby bright northern galaxies. These data have
been supplemented by data and results from two other recent surveys, one by 
\citet{houlv01}, \citet{ulvho01a} and \citet{andet04} and the other by 
\citet{filet00} and \citet{filet04}.
The completion of uniform high resolution radio surveys of the LLAGNs and AGNs 
in the Palomar sample of nearby bright galaxies has yielded
the following main results: \newline
a)~97\% (38 of 39) of the LLAGNs and AGNs (with S$_{\rm VLA}^{\rm 15\,GHz}\,> 2.7$~mJy) 
   investigated at mas resolution with the VLBA 
   have pc-scale nuclei with brightness 
   temperatures $\gtrsim\,10^{6.3}$~K.  Of these nuclei, the ones with the highest 
   nuclear flux densities typically show pc-scale jets. 
   The luminosity, brightness temperature, spectrum, morphology, and 
   variability of the radio emission all argue against an origin in star-formation 
   related processes or as thermal emission. Thus, the nuclear radio emission probably
   originates either in an accretion inflow onto a supermassive black hole or
   from jets launched by this black hole-accretion disk system. The latter
   explanation is supported by the radio morphologies \citep[e.g.][]{falet00,naget02a},
   radio flux variability \citep{naget02a}, and the radio spectral shapes
   \citep{naget01,naget02b};
   \newline
b)~there is no reason to believe that the remaining LLAGNs with compact radio 
   nuclei (investigated at 150~mas resolution) are different from the LLAGNs 
   investigated at mas resolution.
   Thus, at least half of all LINERs and low-luminosity Seyferts probably contain
   accreting black holes. The incidence for transition nuclei is much lower; 
   \newline
c)~compact radio nuclei are preferentially found in massive ellipticals and
   in type~1 nuclei (i.e.  nuclei in which broad \ha\ emission is present). 
   The preferential detection of type~1 nuclei could result: 
   1) from observational selection effects, in which broad \ha\ 
      and radio nuclei have been found in only the more powerful LLAGNs, 
   2) if only the type~1 LLAGNs are bona-fide AGNs, or 
   3) if  the unified scheme applies and the radio emission from type 1
      nuclei is beamed perpendicular to the plane of obscuring material
      and/or type~2 nuclei are free-free absorbed by the obscuring disk in the 
      radio; \newline
d)~the radio luminosity of the compact nucleus is correlated with the galaxy 
   luminosity and the luminosity and width of the nuclear emission lines \citep{naget02a}.
   These trends suggest that we have detected only the brighter LLAGNs, i.e. the 
   true incidence of accreting black holes in LLAGNs is higher than found by our
   survey; \newline
e)~The nuclear radio and nuclear emission-line properties of LLAGNs fall close to the 
   low-luminosity extrapolations of more powerful AGNs, providing further support 
   for a common central engine;  \newline
f)~low-luminosity Seyferts and LINERs share many of the same characteristics in 
   the radio. The transition nuclei detected are those which are the closest, in 
   terms of emission-line diagnostic ratios, to Seyferts and LINERs \citep{naget02a}. 
   Thus at least some transition nuclei are really composite 
   Seyfert/LINER + \hii\ region nuclei, with 
   the nuclear radio luminosity dependent on the Seyfert/LINER component; \newline
g)~investigation of all $\sim\,$50 nearby bright galaxies (most of them 
   LLAGNs) with one radio component relatively unambiguously identified with the
   central engine at sub-parsec resolution, shows that the sub-parsec radio 
   luminosity is 
   correlated with both the estimated mass of the nuclear black hole and the galaxy 
   bulge luminosity.
   The large scatter in radio luminosity at any given black hole mass may be 
   caused by a range of accretion rates (Eddington ratios $\sim$10$^{-1}-10^{-5}$);  \newline
h)~about half of all LLAGNs investigated show significant inter-year variability
   at 15~GHz (2~cm) and 8.4~GHz (3.6~cm) \citep{naget02a};  \newline
i)~the nuclear radio luminosity function at 15~GHz of luminous Seyferts, the Palomar 
   sample, and the local group of galaxies extends over the range 
   10$^{15}-10^{24}$ Watt~Hz$^{-1}$, more than five orders of magnitude larger than 
   previous AGN samples. We find log $\rho\,\propto -0.78\,\times\,$log L$_{\rm 15\,GHz}$
   (where $\rho$ is the space density in Mpc$^{-3}$ mag$^{-1}$), 
   with some evidence for a low luminosity turnover near log L$_{\rm 15\,GHz}\,\sim$ 19; \newline
j)~within the context of jet models, the primary accretion energy output from LLAGNs 
   with compact radio cores is jet kinetic power; this jet power could dominate the radiated 
   bolometric luminosity by factors of $\sim$2 to $>10^2$. These jets, which are 
   energetically more significant than supernovae in the host galaxy, 
   can potentially deposit sufficient energy into the inner parsecs to 
   significantly slow gas flow into the accretion disk; \newline
k) within the context of jet models,
   the mass accretion function for Palomar LLAGNs and AGNs has the 
   form log $\rho$ $\propto -1 \times$ $\dot{\rm M}$. That is, within the Palomar sample,
   the numerous LLAGNs
   with low radio luminosity 
   are together accreting the same mass per year as the fewer LLAGNs and AGNs at higher radio 
   luminosity; 
   \newline
l)~with increasing Eddington ratio, \ledd, LLAGNs in elliptical galaxies
   are increasingly radio-loud as measured by the ratio of radio to optical emission-line
   luminosities; \newline
m) among non-elliptical hosts, LINERs have lower nuclear gas densities, lower Eddington ratios, 
   and are more efficient
   at launching collimated sub-parsec jets, than low-luminosity Seyferts. We speculate that,
   by analogy with  
   Galactic black hole candidates, LINERs are in a `low/hard' state while
   Seyferts are in a `high' state (Sect.~\ref{secdiscussion}); \newline
n)~high resolution radio imaging is an effective and efficient search technique for
   finding low luminosity accreting black holes in LLAGNs. \newline

In short, all evidence points toward the presence of accreting black holes in 
   a large fraction, perhaps all, of LLAGNs. Compact radio jets are an 
   energetically important product of accretion in this low luminosity regime. 
   The radio luminosity function and the jet kinetic powers of these LLAGNs together show
   that jets are the dominant form of power output from the nuclear accretion 
   and suggest that the 
   jets may input a significant amount of energy into the ISM of their host galaxies.
   
\acknowledgements
This work was partly funded by the Dutch research organization
NWO, through a VENI grant to NN, and by NASA through grants NAG513065
and NAG513557 to the University of Maryland. Parts of this work were
completed while NN held a fellowship at Arcetri Observatory, Italy.

\section{Appendix: Absorption-line nuclei in the Palomar Sample}

The Palomar spectroscopic survey did not detect emission lines in 53 nuclei of the 
Palomar Sample; upper limits to the H$\alpha$ luminosity in these nuclei
are listed in \citet{hoet03a}{\footnote{\citet{hoet03a} list 54 nuclei 
with H$\alpha$ luminosity upper limits; of these, the nucleus of NGC~4494 
shows other emission lines and was classified
as a (highly uncertain) `LINER 2' by \citet{hoet97a}.}}
We have searched the literature and VLA archive for high resolution radio observations
of these 53 nuclei. The results are listed in Table~\ref{tabpalabs}. Only two of the
nuclei have AGN-related radio fluxes significantly greater than $\sim$1~mJy. 
NGC~507 (also known as B2~0120+33) has about 100~mJy at 1.4~GHz in larger scale radio lobes but 
a very weak ($\sim$1.4~mJy at 5~GHz) radio nucleus \citep{gioet88}; 
it may be that the AGN is now switched
off or in a very low state, which could account for the lack of emission-lines.
NGC~4649 has a 1.4~GHz flux of 18~mJy which is distributed in a core plus twin jet
structure \citep{stawar86}. 
Another three (possibly 5) of the 53 nuclei have weak 
($\sim$1~mJy) radio nuclei, potentially AGN-related, at 1--5{\arcsec} resolution
(Table~\ref{tabpalabs}).

 \begin{figure*}
  \includegraphics[bb=10 220 586 632,width=\textwidth,clip]{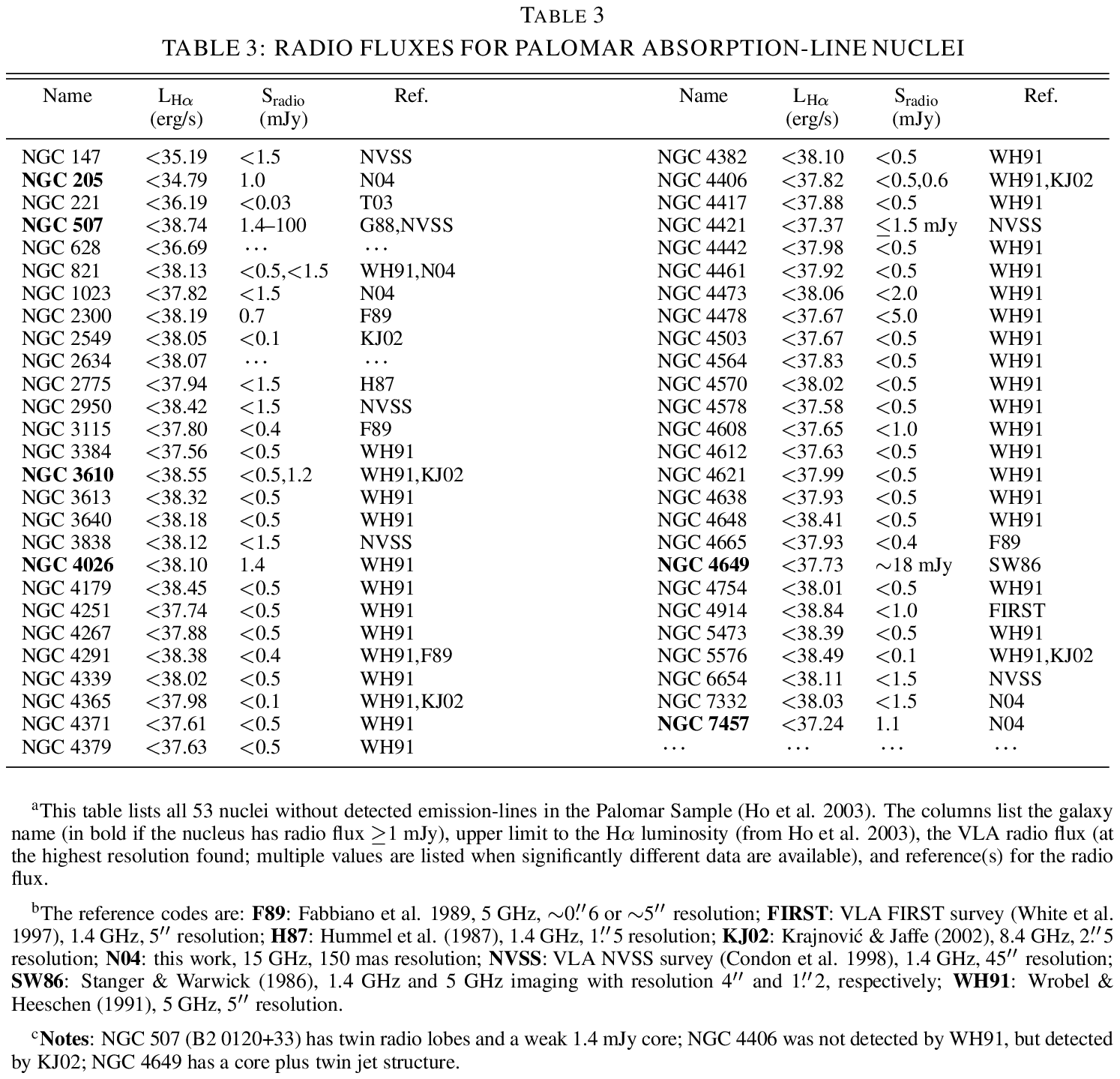}
 \end{figure*}

 \begin{table}
     \dummytable\label{tabvla}
 \end{table}
 \begin{table}
     \dummytable\label{tabvlba}
 \end{table}
 \begin{table}
     \dummytable\label{tabpalabs}
 \end{table}
 \begin{table}
     \dummytable\label{tabrlf}
 \end{table}
 
\clearpage
 \begin{figure*}
   \includegraphics[width=\textwidth,clip]{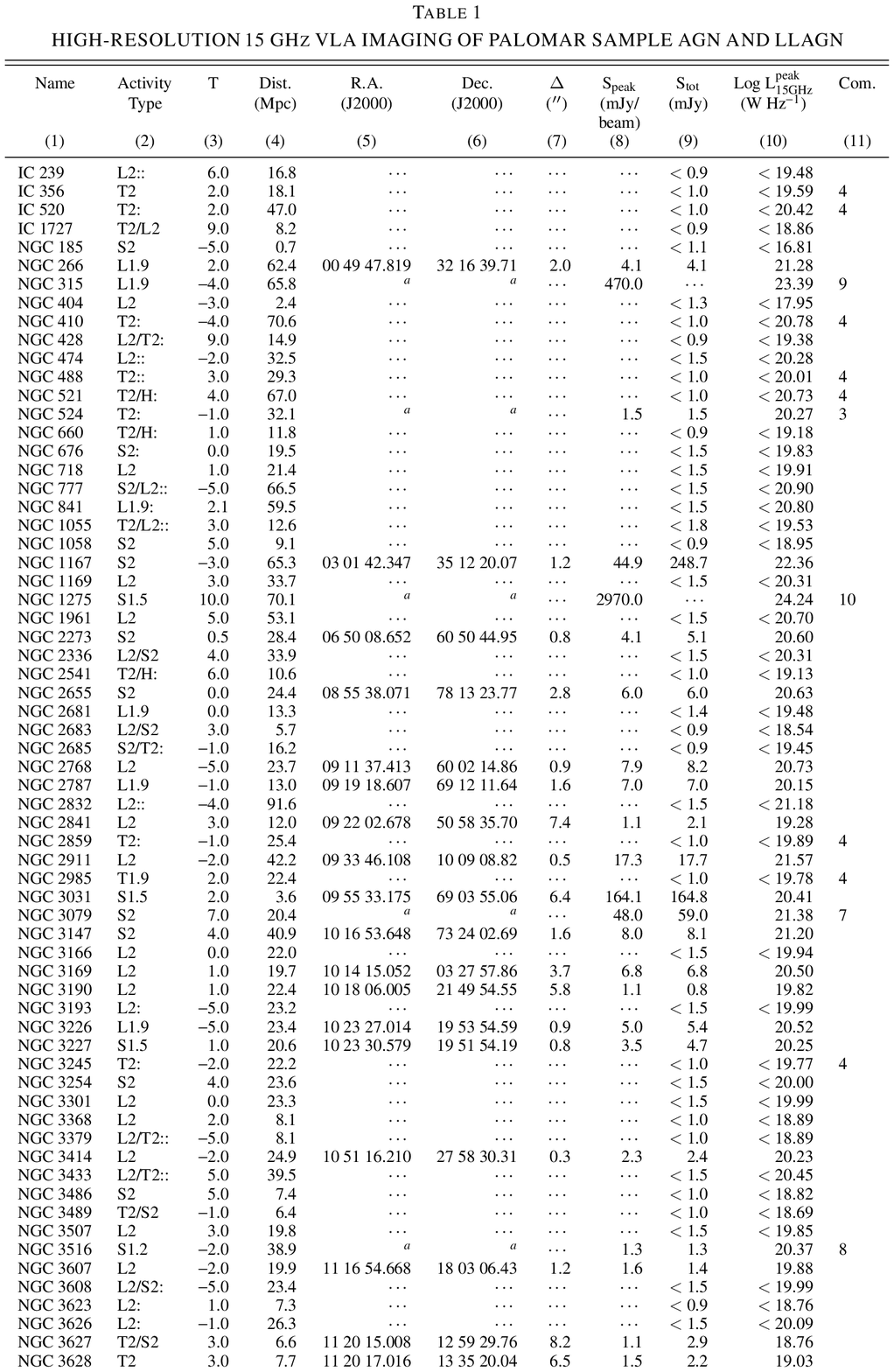} 
 \end{figure*}
 \begin{figure*}
   \includegraphics[width=\textwidth,clip]{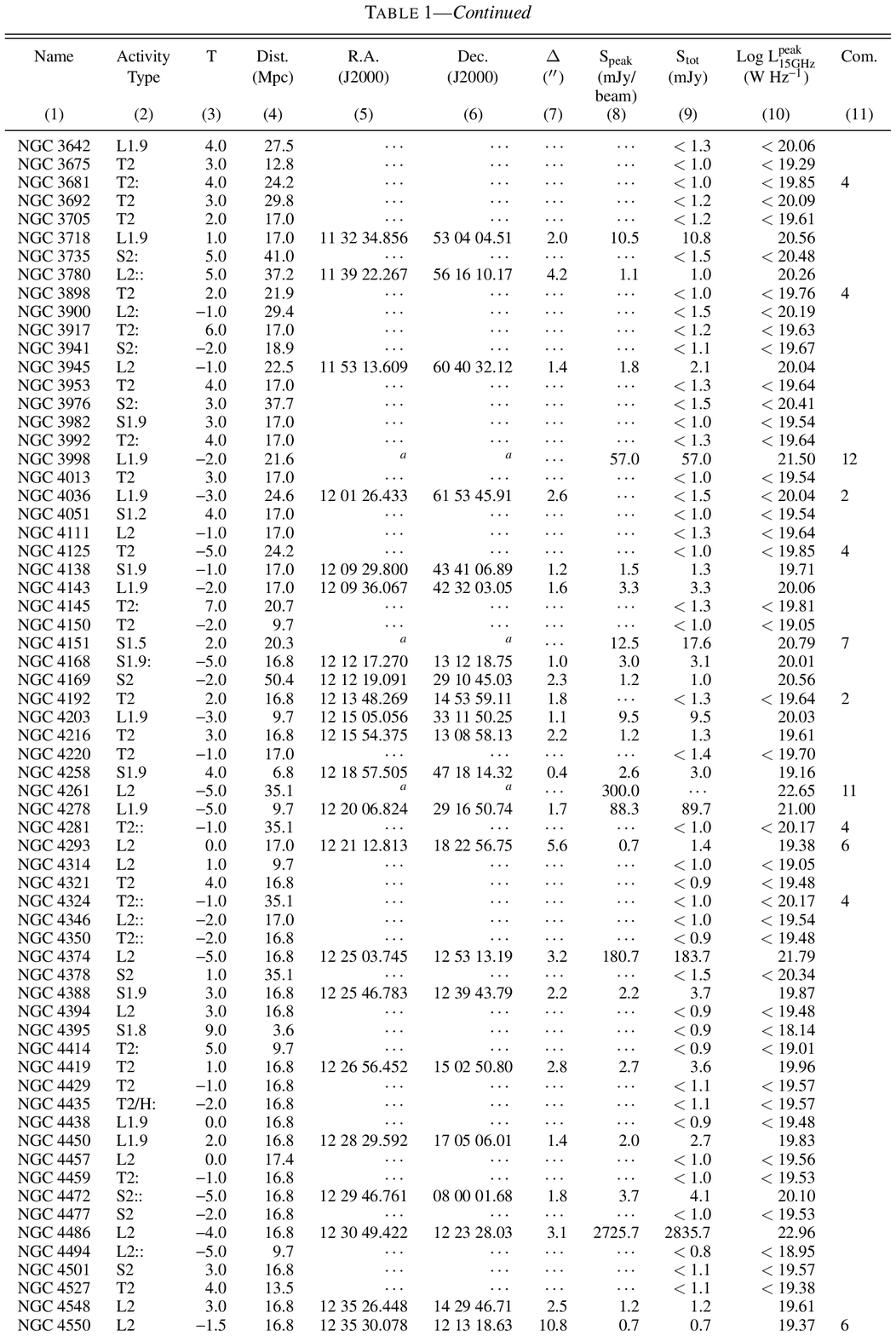} 
 \end{figure*}
 \begin{figure*}
   \includegraphics[width=\textwidth,clip]{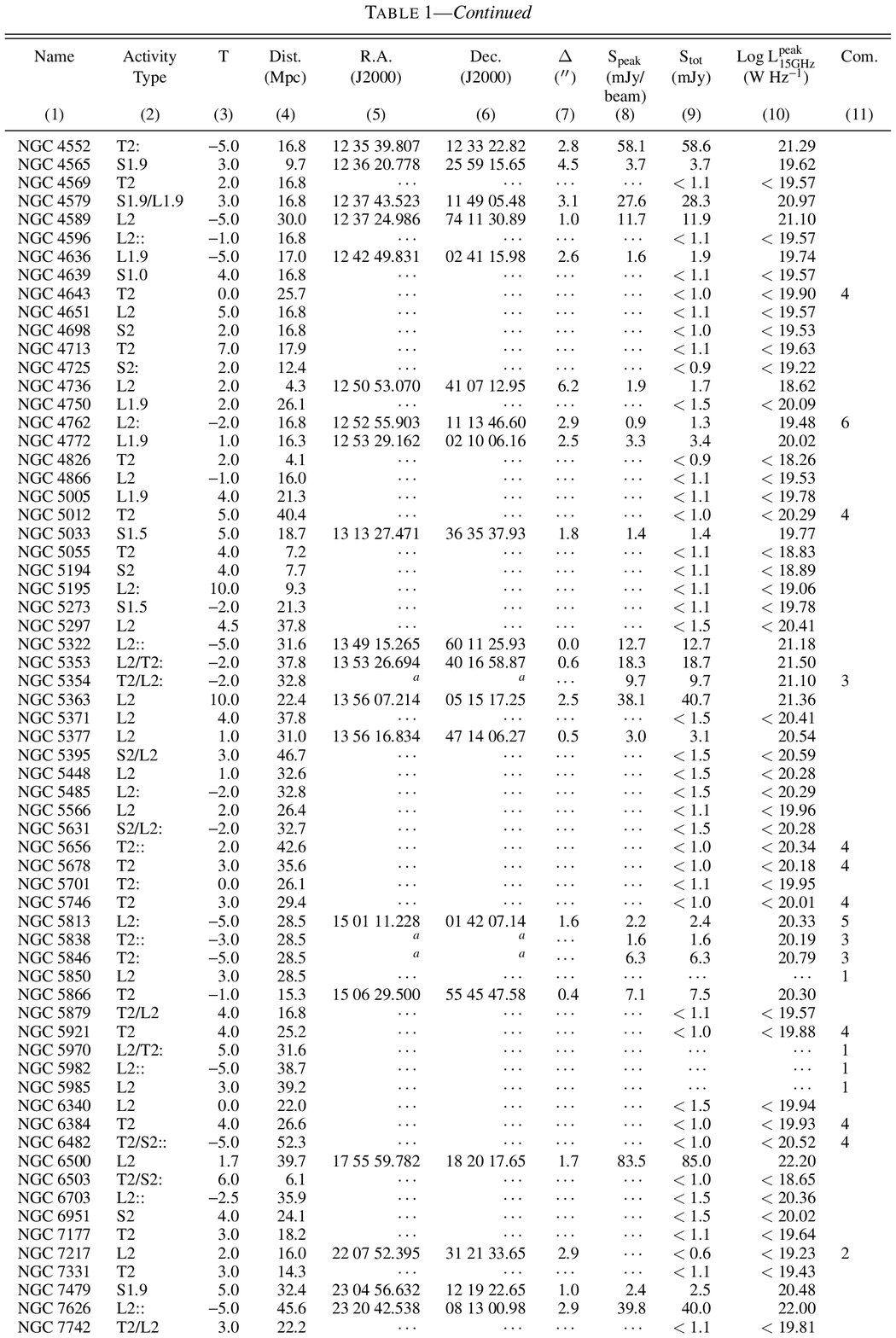} 
 \end{figure*}
 \begin{figure*}
   \includegraphics[width=\textwidth,clip]{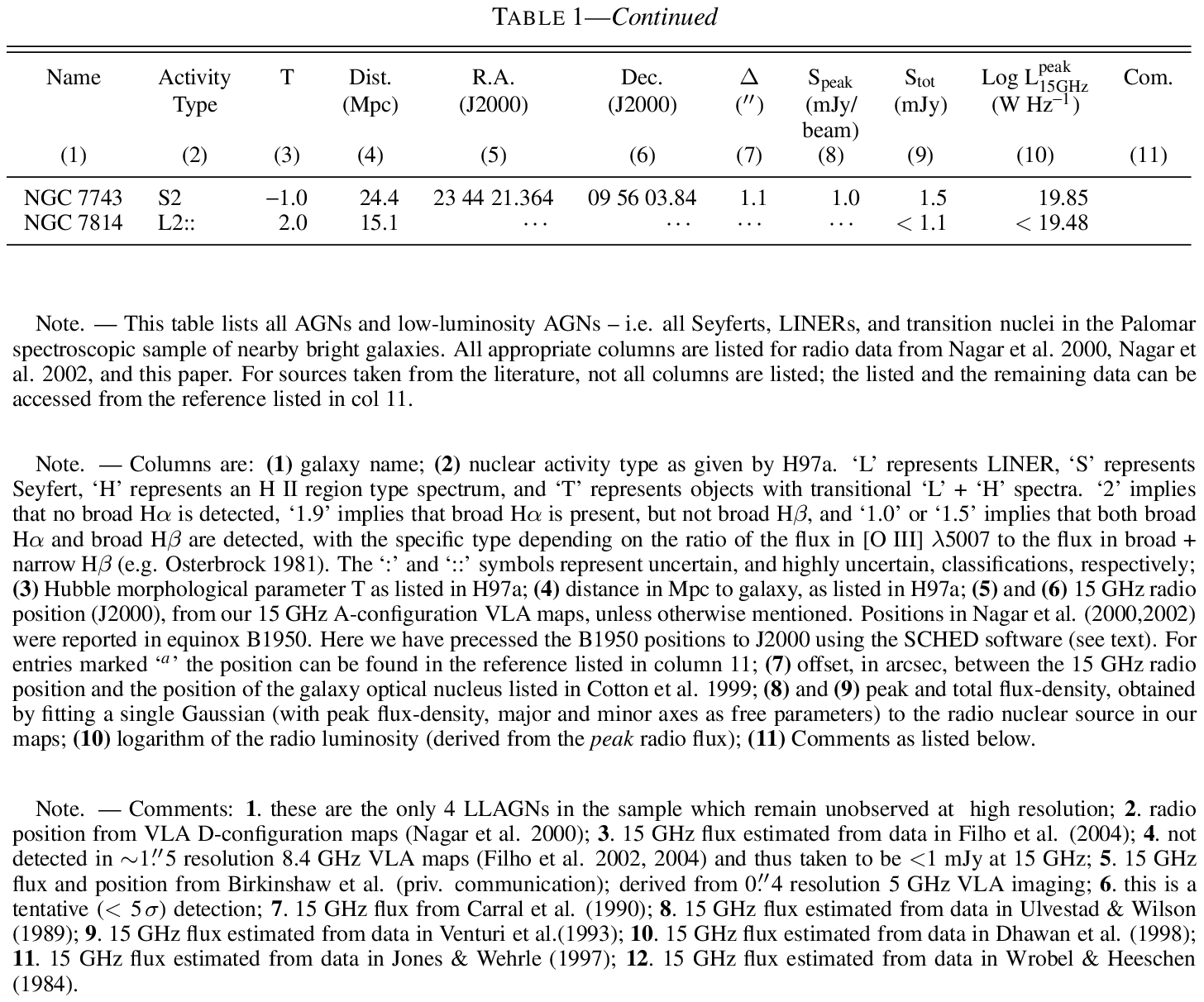} 
 \end{figure*}

 \begin{figure*}
  \includegraphics[bb=20 50 577 800,clip,width=7.2in]{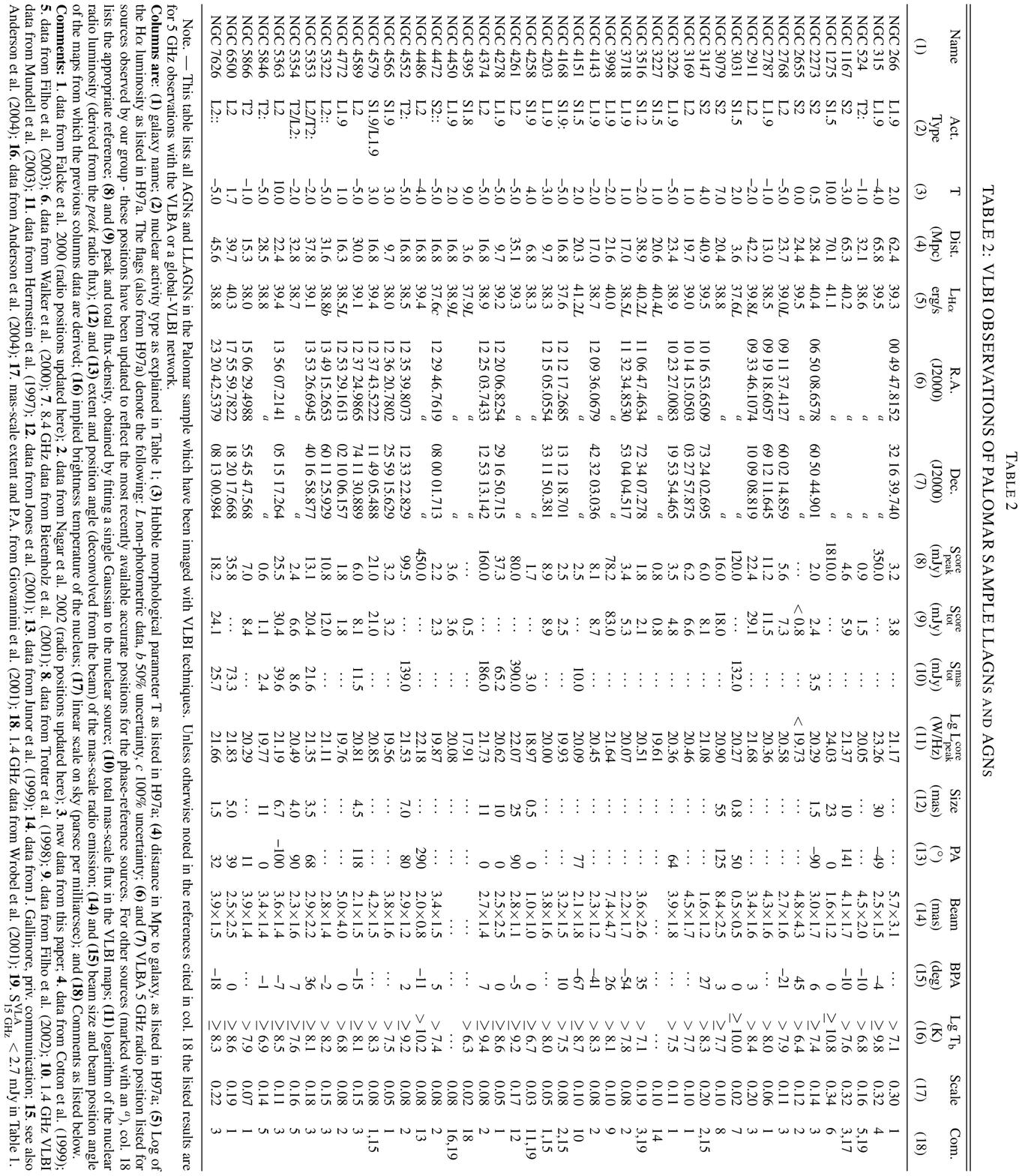}
 \end{figure*}


\begin{thebibliography}{dummy}
\bibitem[Alonso-Herrero et al.(1999)]{aloet99}
Alonso-Herrero, A., Rieke, M. J., Rieke, G. H., \& Shields, J. C.
2000, ApJ, 530, 688       


\bibitem[Anderson, Ulvestad, \& Ho(2004)]{andet04} 
Anderson, J.~M., Ulvestad, J.~S., \& Ho, L.~C.\ 2004, \apj, 603, 42 

\bibitem[Antonucci(1993)]{ant93}
Antonucci, R. R. J. 1993, \araa, 31, 473 


\bibitem[Barth et al.(1999)]{baret99}
Barth, A. J., Filippenko, A. V., Moran, E. C.
1999, \apj, 525, 673

\bibitem[Barth et al.(1998)]{baret98}
Barth, A. J., Ho, L. C., Filippenko, A. V., \&
Sargent, W. L. W.  1998, \apj, 496, 133




\bibitem[Beasley et al.(2002)]{beaet02}
Beasley, A.~J., Gordon, D., Peck, A.~B., Petrov, L., 
MacMillan, D.~S., Fomalont, E.~B., \& Ma, C.\ 2002, \apjs, 141, 13 

\bibitem[Begelman, Blandford, \& Rees(1984)]{beget84}
Begelman, M. C., Blandford, R. D., \& Rees, M. J. 1984,
Rev. Mod. Phys., 56, 255  

\bibitem[Bietenholz et al.(2000)]{bieet00}
Bietenholz, M. F., Bartel, N., Rupen, M. P., 2000, \apj, 532, 895               
 
 \bibitem[Binney(2004a)]{bin04a}
 Binney, J. 2004a, to appear in \mnras (astro-ph/0308171)

 \bibitem[Binney(2004b)]{bin04b}
 Binney, J. 2004b, to appear in Phil Trans Roy Soc (astro-ph/0407238)

 \bibitem[Binney \& Tremaine(1987)]{bintre87} 
 Binney, J.~\& Tremaine, S.\ 1987, `Galactic Dynamics', 
 Princeton, NJ, Princeton University Press, 1987, 747 p.  


\bibitem[Blandford(1993)]{bla93}
Blandford, R. D. 1993, in Astrophysical Jets, ed.
D. Burgarella, M. Livio, \& C. P. O'Dea, (Cambridge:
Cambridge Univ. Press), 15


\bibitem[Braatz et al.(1997)]{braet97}
Braatz, J., Wilson, A. S., \& Henkel, C. 1997, \apjs, 110, 321    

\bibitem[Carral, Turner, \& Ho(1990)]{caret90} Carral, P.,
Turner, J.~L., \& Ho, P.~T.~P.\ 1990, \apj, 362, 434 

 \bibitem[Celotti \& Fabian(1993)]{celfab93} 
 Celotti, A.~\& Fabian, A.~C.\ 1993, \mnras, 264, 228 


\bibitem[Chatterjee et al.(2004)]{chaet04} 
Chatterjee, S., Cordes, J.~M., Vlemmings, W.~H.~T., Arzoumanian, Z., 
Goss, W.~M., \& Lazio, T.~J.~W.\ 2004, \apj, 604, 339 

\bibitem[Colina et al.(2001)]{colet01}
Colina, L., Alberdi, A., Torrelles, J. M.,
Panagia, N., \& Wilson, A. S. 2001, \apjl, 553, L19

\bibitem[Condon(1989)]{con89}
Condon, J.~J.\ 1989, \apj, 338, 13 

 \bibitem[Condon et al.(1998)]{conet98}
 Condon, J. J., Cotton, W. D., Greisen, E. W., Yin, Q. F., Perley, R. A., 
 Taylor, G. B., \& Broderick, J. J. 1998, \aj, 115, 1693


\bibitem[Condon et al.(1991)]{conet91}
Condon, J. J., Huang, Z.-P., Yin, Q. F., \& Thuan, T. X. 1991
\apj, 378, 65 


 \bibitem[Cotton et al.(1999a)]{cotet99a} 
 Cotton, W.~D., Condon, J.~J., \& Arbizzani, E.\ 1999, \apjs, 125, 409 

 \bibitem[Cotton et al.(1999b)]{cotet99b}
 Cotton, W.~D., Feretti, L., Giovannini, G., Lara, L., \& Venturi, T.\
 1999, \apj, 519, 108


 \bibitem[Crane et al.(1993)]{craet93}
 Crane, P. C., Cowan, J. J., Dickel, J. R., \& Roberts, D. A. 1993,
 \apjl, 417, L61



\bibitem[Dhawan, Kellerman, \& Romney(1998)]{dhaet98} Dhawan,
V., Kellerman, K.~I., \& Romney, J.~D.\ 1998, \apjl, 498, L111   

 \bibitem[Di Matteo et al.(2001)]{dimet01}
 Di Matteo, T., Carilli, C. L., \& Fabian, A. C.  2001, \apj, 547, 731
 

\bibitem[Dopita \& Sutherland(1995)]{dopsut95}
Dopita, M. A., \& Sutherland, R. S. 1995, \apj, 455, 468



\bibitem[Emsellem et al.(1999)]{emset99} 
Emsellem, E., Dejonghe, H., \& Bacon, R.\ 1999, \mnras, 303, 495 


\bibitem[Fabbiano, Gioia \& Trinchieri(1989)]{fabet89}
Fabbiano, G., Gioia, I. M., \& Trinchieri, G. 1989, \apj, 374, 127

\bibitem[Falcke \& Biermann(1999)]{falbie99}
Falcke, H. \& Biermann, P. L. 1999, \aap, 342, 49    


\bibitem[Falcke, K{\" o}rding, \& Markoff(2004)]{falet04}
Falcke, H., K{\" o}rding, E., \& Markoff, S.\ 2004, \aap, 414, 895 

\bibitem[Falcke et al.(2000)]{falet00}
Falcke, H., Nagar, N. M., Wilson, A. S., \& Ulvestad, J. S. 2000, \apj,
542, 197  (Paper II)



\bibitem[Fender \& Belloni(2004)]{fenbel04} 
Fender, R.~\& Belloni, T.\ 2004, \araa, 42, 317 

\bibitem[Fender, Gallo, \& Jonker(2003)]{fenet03} 
Fender, R.~P., Gallo, E., \& Jonker, P.~G.\ 2003, \mnras, 343, L99 


 \bibitem[Ferrarese \& Merritt(2000)]{fermer00}
 Ferrarese, L, \& Merritt, D. 2000, \apjl, 539, L9

 \bibitem[Filho et al.(2000)]{filet00}
 Filho, M. E., Barthel, P. D., \& Ho, L. C. 2000, \apjs, 129, 93
 
 \bibitem[Filho, Barthel, \& Ho(2002)]{filet02}
 Filho, M.~E., Barthel, P.~D., \& Ho, L.~C.\ 2002, \aap, 385, 425

 \bibitem[Filho et al.(2004)]{filet04} 
 Filho, M.~E., Fraternali, 
 F., Markoff, S., Nagar, N.~M., Barthel, P.~D., Ho, L.~C., \& Yuan, F.\ 
 2004, \aap, 418, 429 

\bibitem[Filippenko \& Ho(2003)]{filho03} 
Filippenko, A.~V., \& Ho, L.~C.\ 2003, \apjl, 588, L13 

\bibitem[Filippenko \& Terlevich(1992)]{filter92}
Filippenko, A. V., \& Terlevich, R. 1992, \apjl, 397, L79

\bibitem[Fomalont et al.(2000)]{fomet00} 
Fomalont, E.~B., Frey, S., Paragi, Z., Gurvits, L.~I., Scott, W.~K., 
Taylor, A.~R., Edwards, P.~G., \& Hirabayashi, H.\ 2000, \apjs, 
131, 95

\bibitem[Fosbury et al.(1978)]{foset78}
Fosbury, R. A. E., Mebold, U., Goss, W. M., \& Dopita, M. A.
1978, \mnras, 183, 549                

\bibitem[Franceschini et al.(1998)]{fraet98}
Franceschini, A., Vercellone, S., \& Fabian, A. C. 1998, \mnras, 297, 817

\bibitem[Frank, King, \& Raine(1995)]{fraet95}
Frank, J., King, A., \& Raine, D. 1995, in Accretion
Power in Astrophysics, 2nd edition,
(Cambridge: Cambridge Univ. Press)  




 \bibitem[Gebhardt et al.(2000)]{gebet00}
 Gebhardt, K., et al.\ 2000, \apjl, 539, L13

 \bibitem[Gebhardt et al.(2003)]{gebet03} 
 Gebhardt, K., et al.\ 2003, \apj, 583, 92 



 \bibitem[Giovannini et al.(1988)]{gioet88}
 Giovannini, G., Feretti, L., Gregorini, L., \& Parma, P.\ 1988, \aap, 199, 73 

 \bibitem[Giovannini et al.(2001)]{gioet01} 
 Giovannini, G., Cotton, W.~D., Feretti, L., Lara, L., \& Venturi, 
 T.\ 2001, \apj, 552, 508


\bibitem[Heckman(1980)]{hec80}
Heckman, T. M. 1980, \aap, 87, 152

\bibitem[Heckman, Crane, \& Balick(1980)]{hecet80}
Heckman, T.~M., Crane, P.~C., \& Balick, B.\ 1980, \aaps, 40, 295 

\bibitem[Heckman et al.(2004)]{hecet04} 
Heckman, T.~M., Kauffmann, G., Brinchmann, J., Charlot, S., Tremonti, C., 
\& White, S.~D.~M.\ 2004, \apj, 613, 109 



 \bibitem[Herrnstein et al.(1997)]{heret97}
 Herrnstein, J. R., Moran, J. M., Greenhill, L. J., Diamond, P. J., 
 Miyoshi, M., Nakai, N., \& Inoue, M. 1997, \apjl, 475, L17

\bibitem[Ho(2002)]{ho02} Ho, L.~C.\ 2002, \apj, 564, 120 

\bibitem[Ho(1999)]{ho99}
Ho, L. C. 1999, \apj, 516, 672

\bibitem[Ho et al.(2001)]{hoet01}
Ho, L.~C.~et al.\ 2001, \apjl, 549, L51


\bibitem[Ho et al.(1995)]{hoet95}
Ho, L. C., Filippenko, A. V., \& Sargent, W. L. W. 1995, \apjs, 98, 477


\bibitem[Ho et al.(1997a)]{hoet97a}
Ho, L. C., Filippenko, A. V., \& Sargent, W. L. W. 1997a, 
\apjs, 112, 315 


\bibitem[Ho, Filippenko, \& Sargent(2003a)]{hoet03a} 
Ho, L.~C., Filippenko, A.~V., \& Sargent, W.~L.~W.\ 2003, \apj, 583, 159 

\bibitem[Ho et al.(1997b)]{hoet97b}
Ho, L. C., Filippenko, A. V., \& Sargent, W. L. W., \& Peng, C. Y.
1997b, \apjs, 112, 391 

\bibitem[Ho et al.(2003b)]{hoet03b}
Ho, L.~C., Terashima, Y., \& Ulvestad, J.~S.\ 2003, \apj, 589, 783 

\bibitem[Ho \& Ulvestad(2001)]{houlv01} 
Ho, L.~C.~\& Ulvestad, J.~S.\ 2001, \apjs, 133, 77 



 \bibitem[Hummel(1980)]{hum80} 
 Hummel, E.\ 1980, \aaps, 41, 151 

 \bibitem[Hummel et al.(1982)]{humet82} 
 Hummel, E., Fanti, C., Parma, P., \& Schilizzi, R.~T.\ 1982, \aap, 114, 400 

 \bibitem[Hummel et al.(1987)]{humet87}
 Hummel, E., van der Hulst, J.~M., Keel, W.~C., \& Kennicutt, 
 R.~C.\ 1987, \aaps, 70, 517 


\bibitem[Jones \& Wehrle(1997)]{jonweh97}
Jones, D. L., \& Wehrle, A. E. 1997, \apj, 484, 186

 \bibitem[Jones et al.(2001)]{jonet01}
 Jones, D.~L., Wehrle, A.~E., Piner, B.~G., \& Meier, D.~L.\
 2001, \apj, 553, 968


\bibitem[Jones, Terzian, \& Sramek(1981)]{jonet81} Jones, 
D.~L., Terzian, Y., \& Sramek, R.~A.\ 1981, \apj, 246, 28 

\bibitem[Jones et al.(1984)]{jonet84}
Jones, D. L., Wrobel, J. M., \& Shaffer, D. B. 1984, \apj, 276, 480  

\bibitem[Junor \& Biretta(1995)]{junbir95}
Junor, W., \& Biretta, J. A. 1995, \aj, 109, 500


 \bibitem[Karachentsev, Makarov, \& Huchtmeier(1999)]{karet99} 
 Karachentsev, I.~D., Makarov, D.~I., \& Huchtmeier, W.~K.\ 1999, \aaps, 
 139, 97 

   
 \bibitem[Kellermann et al.(1998)]{kelet98} Kellermann, K.~I., 
 Vermeulen, R.~C., Zensus, J.~A., \& Cohen, M.~H.\ 1998, \aj, 115, 1295 

\bibitem[Koski \& Osterbrock(1976)]{kosost76}
Koski, A. T., \& Osterbrock, D. E. 1976, \apjl, 203, L49       

\bibitem[Krajnovi{\' c} \& Jaffe(2002)]{krajaf02}
Krajnovi{\' c}, D.~\& Jaffe, W.\ 2002, \aap, 390, 423 

 \bibitem[Krichbaum et al.(1998)]{kriet98}
 Krichbaum, T. B., et al. 1998, \aap, 335, L106

\bibitem[Kukula et al.(1999)]{kuket99}
Kukula, M.~J., Ghosh, T., Pedlar, A., \& Schilizzi, R.~T.\ 1999, \apj, 518, 117

\bibitem[Kukula et al.(1995)]{kuket95} 
Kukula, M. J., Pedlar, A., Baum, S. A., \& O'Dea, C. P.  
1995, \mnras, 276, 1262



 \bibitem[Laurent-Muehleisen et al.(1997)]{lauet97}
 Laurent-Muehleisen, S. A., Kollgaard, R. I., Ryan, P. J.,
 Feigelson, E. D., Brinkmann, W., \& Siebert, J.
 1997, \aaps, 122, 235

 \bibitem[Lavalley et al.(1992)]{lavet92}
 Lavalley, M., Isobe, T., \& Feigelson, E. 1992, in Astronomical
 Data Analysis Software and Systems I,
 ed. D. Worrall, C. Biemesderfer \& J. Barnes, (San Francisco: ASP),
 Vol. 25, 245 (ASURV)


 \bibitem[Lonsdale, Lonsdale, \& Smith(1992)]{lonet92}
 Lonsdale, C.~J., Lonsdale, C.~J., \& Smith, H.~E.\ 1992, \apj, 391, 629

 \bibitem[Lonsdale, Smith, \& Lonsdale(1993)]{lonet93} 
 Lonsdale, C.~J., Smith, H.~J., \& Lonsdale, C.~J.\ 1993, \apjl, 405, L9

\bibitem[Lovelace \& Romanova(1996)]{lovrom96}
Lovelace, R. V. E., \& Romanova, M. M. 1996, in Energy Transport in
Radio Galaxies, ed. P. E. Hardee, A. H. Bridle, \& J. A. Zensus
(San Francisco: ASP), Vol. 100, 25  




\bibitem[Maoz et al.(1995)]{maoet95}
Maoz, D., Filippenko, A. V., Ho, L. C., Rix, H.-W., Bahcall, J. N.,
Schneider, D. P., \& Macchetto, F. D. 1995, \apj, 440, 91        

\bibitem[Maoz et al.(1998)]{maoet98} Maoz, D., Koratkar, A., 
Shields, J.~C., Ho, L.~C., Filippenko, A.~V., \& Sternberg, A.\ 1998, \aj, 
116, 55 

\bibitem[Marconi \& Hunt(2003)]{marhun03} 
Marconi, A.~\& Hunt, L.~K.\ 2003, \apjl, 589, L21 


\bibitem[Merloni, Heinz, \& di Matteo(2003)]{meret03} 
Merloni, A., Heinz, S., \& di Matteo, T.\ 2003, \mnras, 345, 1057 

 \bibitem[Merritt \& Ferrarese(2001)]{merfer01} 
 Merritt, D.~\& Ferrarese, L.\ 2001, \mnras, 320, L30

 \bibitem[Meurs \& Wilson(1984)]{meuwil84} 
 Meurs, E.~J.~A.~\& Wilson, A.~S.\ 1984, \aap, 136, 206 


 \bibitem[Middelberg et al.(2004)]{midet04}
 Middelberg, E., et al.\ 2004, \aap, 417, 925 

 \bibitem[Mundell et al.(2000)]{munet00}
 Mundell, C.~G., Wilson, A.~S., Ulvestad, J.~S., \& Roy, A.~L.\ 
 2000, \apj, 529, 816

 \bibitem[Mundell et al.(2003)]{munet03}
 Mundell, C.~G., Wrobel, J.~M., Pedlar, A., \& Gallimore, J.~F.\
 2003, \apj, 583, 192

\bibitem[Nagar(2003)]{nag03}
Nagar, N. M.  2003, to appear in the proceedings of `Multiwavelength AGN Surveys', 
Cozumel, December 2003, eds. R. Maiolino \& R. Mujica (World Scientific)

\bibitem[Nagar et al.(2000)]{naget00}
Nagar, N. M., Falcke, H., Wilson, A. S., \& Ho, L. C.
2000, \apj, 542, 186 (Paper~I) 

\bibitem[Nagar et al.(2002a)]{naget02a}
Nagar, N.~M., Falcke, H., Wilson, A.~S., \& Ulvestad, J.~S.\ 2002a, \aap, 392, 53 

\bibitem[Nagar et al.(2001)]{naget01} 
Nagar, N.~M., Wilson, A.~S., \& Falcke, H.\ 2001, \apjl, 559, L87 

\bibitem[Nagar et al.(2002b)]{naget02b}
Nagar, N.~M., Wilson, A.~S., Falcke, H., Ulvestad, J.~S., \& Mundell, C.~G.\ 2002b, 
ASP Conf.~Ser.~258: Issues in Unification of Active Galactic Nuclei, 171 

\bibitem[Nagar et al.(1999)]{naget99}
Nagar, N. M., Wilson, A. S., Mulchaey, J. S., \& Gallimore, J. F. 
1999, \apjs, 120, 209

\bibitem[Napier et al.(1994)]{napet94}
Napier, P. J., Bagri, D. S., Clark, B. G., Rogers, A. E. E.,
Romney, J. D., Thompson, A. R., \& Walker, R. C. 1994, Proc.
IEEE, 82, 658

\bibitem[Narayan et al.(1998)]{naret98}
Narayan, R., Mahadevan, R., \& Quataert, E. 1998, in
The Theory of Black Hole Accretion Discs, ed.  M. A. Abramowicz,
G. Bj\"{o}rnsson, \& J. E. Pringle (Cambridge: Cambridge Univ. Press), 148

\bibitem[Narayan et al.(2000)]{naret00} Narayan, R., Igumenshchev, I. V., 
\& Abramowicz, M. A. 2000, \apj, 539, 798         


\bibitem[Neff \& Ulvestad(2000)]{nefulv00} 
Neff, S. G., \& Ulvestad, J. S. 2000, \aj, 120, 670



\bibitem[Osterbrock(1981)]{ost81}
Osterbrock, D. E. 1981, \apj, 249, 462

\bibitem[Osterbrock(1989)]{ost89}
Osterbrock, D. E. 1989, Astrophysics of Gaseous Nebulae and Active
Galactic Nuclei (Mill Valley, CA: Univ. Sci. Books)

\bibitem[Ostriker \& Ciotti(2004)]{ostcio04}
 Ostriker, J. P., \& Ciotti, L. 2004, 
 to appear in Phil Trans Roy Soc (astro-ph/0407234)

 \bibitem[Owen, Eilek, \& Kassim(2000)]{oweet00} 
 Owen, F.~N., Eilek, J.~A., \& Kassim, N.~E.\ 2000, \apj, 543, 611 


 \bibitem[Pellegrini \& Ciotti(1998)]{pelcio98} 
 Pellegrini, S.~\& Ciotti, L.\ 1998, \aap, 333, 433 


\bibitem[Pringle(1993)]{pri93}
Pringle, J. E. 1993, in Astrophysical Jets, ed. D. Burgerella, M. Livio,
\& C. P. O'Dea, (Cambridge: Cambridge Univ. Press), 1 



 \bibitem[Rawlings \& Saunders(1991)]{rawsau91} 
 Rawlings, S.~\& Saunders, R.\ 1991, \nat, 349, 138 

\bibitem[Rees et al.(1982)]{reeet82} 
Rees, M.~J., Phinney, E.~S., Begelman, M.~C., \& Blandford, R.~D.\ 
1982, \nat, 295, 1

 \bibitem[Richstone et al.(1998)]{ricet98}
 Richstone, D., et al. 1998, \nat, 395A, 14




\bibitem[Sadler et al.(1989)]{sadet89}
Sadler, E. M., Jenkins, C. R., \& Kotanyi, C. G. 1989,
\mnras, 240, 591

\bibitem[Sadler et al.(1995)]{sadet95}
Sadler, E. M., Slee, O. B., Reynolds, J. E., \& Roy, A. L.
1995, \mnras, 276, 1373

\bibitem[Sandage, Tammann, \& Yahil(1979)]{sanet79} 
Sandage, A., Tammann, G.~A., \& Yahil, A.\ 1979, \apj, 232, 352 

\bibitem[Sanghera et al.(1995)]{sanet95}
Sanghera, H. S., Saikia, D. J., Luedke, E., Spencer, R. E., Foulsham, P. A.,
Akujor, C. E., \& Tzioumis, A. K. 1995, \aap, 295, 629


\bibitem[Schilizzi et al.(1983)]{schet83} 
Schilizzi, R.~T., Fanti, C., Fanti, R., \& Parma, P.\ 1983, \aap, 126, 412 




\bibitem[Shields(1992)]{shi92}
Shields, J. C. 1992, \apj, 399, L27 



 \bibitem[Sjouwerman et al.(in prep)]{sjoet04}
 Sjouwerman, L., Mioduszewski, A., \& Greisen, T. 2004, in preparation

\bibitem[Slee et al.(1994)]{sleet94}
Slee, O. B., Sadler, E. M., Reynolds, J. E., \& Ekers, R. D.
1994, \mnras, 269, 928

\bibitem[Smith et al.(1998)]{smiet98}
Smith, H. E., Lonsdale, C. J., \& Lonsdale, C. J. 1998, \apj, 492, 137

 \bibitem[Soltan(1982)]{sol82} 
 Soltan, A.\ 1982, \mnras, 200, 115 


 \bibitem[Stanger \& Warwick(1986)]{stawar86} 
 Stanger, V.~J.~\& Warwick, R.~S.\ 1986, \mnras, 220, 363 



 \bibitem[Stoughton et al.(2002)]{stoet02} 
 Stoughton, C., et al.\ 2002, \aj, 123, 485 


\bibitem[Terashima, Ho, \& Ptak(2000)]{teret00} 
Terashima, Y., Ho, L.\ C., \& Ptak, A.\ F.\ 2000, \apj, 539, 161 

 \bibitem[Terashima \& Wilson(2003)]{terwil03}
 Terashima, Y.~\& Wilson, A.~S.\ 2003, \apj, 583, 145

\bibitem[Terlevich \& Melnick(1985)]{termel85}
Terlevich, R., \& Melnick J. 1985, \mnras, 213, 841

\bibitem[Terlevich et al.(1992)]{teret92}
Terlevich, R., Tenorio-Tagle, G., Franco, J., \& Melnick, J.
1995, \mnras, 272, 198     

\bibitem[Thompson et al.(1980)]{thoet80} Thompson, A. R., Clark, B. G.,
Wade C. M., \& Napier, P. J. 1980, \apjs, 44, 151     

 \bibitem[Tremaine et al.(2002)]{treet02} 
 Tremaine, S., et al.\ 2002, \apj, 574, 740 

 \bibitem[Trotter et al.(1998)]{troet98} 
 Trotter, A.~S., Greenhill, L.~J., Moran, J.~M., Reid, M.~J., 
 Irwin, J.~A., \& Lo, K.\ 1998, \apj, 495, 740 


\bibitem[Ulvestad \& Ho(2001a)]{ulvho01a} 
Ulvestad, J.~S.~\& Ho, L.~C.\ 2001, \apj, 558, 561

\bibitem[Ulvestad \& Ho(2001b)]{ulvho01b} 
Ulvestad, J.~S.~\& Ho, L.~C.\ 2001, \apjl, 562, L133

 \bibitem[Ulvestad et al.(1998)]{ulvet98}
 Ulvestad, J.~S., Roy, A.~L., Colbert, E.~J.~M., \& Wilson, A.~S.\
 1998, \apj, 496, 196



\bibitem[Ulvestad \& Wilson(1989)]{ulvwil89}
Ulvestad, J. S., \& Wilson, A. S. 1989, \apj, 343, 659    

 \bibitem[Ulvestad et al.(1999)]{ulvet99}
 Ulvestad, J. S., Wrobel, J. M., Roy, A. L., Wilson, A. S.,
 Falcke, H., \& Krichbaum, T. P. 1999, \apjl, 517, L81

\bibitem[Woo \& Urry(2002)]{woourr02} 
Woo, J., \& Urry, C.~M.\ 2002, \apj, 579, 530 




\bibitem[Venturi et al.(1993)]{venet93} 
Venturi, T., Giovannini, G., 
Feretti, L., Comoretto, G., \& Wehrle, A.~E.\ 1993, \apj, 408, 81 

 \bibitem[van den Bergh \& McClure(1994)]{vanmcc94} 
 van den Bergh, S.~\& McClure, R.~D.\ 1994, \apj, 425, 205 

\bibitem[Verdoes Kleijn et al.(2002)]{veret02}
Verdoes Kleijn, G.~A., Baum, S.~A., de
Zeeuw, P.~T., \& O'Dea, C.~P.\ 2002, \aj, 123, 1334

\bibitem[Vila et al.(1990)]{vilet90}
Vila, M. B., Pedlar, A., Davies, R. D., Hummel, E. \& Axon,
D. J. 1990, \mnras, 242, 379

 \bibitem[Walker et al.(2000)]{walet00}
 Walker, R.~C., Dhawan, V., Romney, J.~D., Kellermann, K.~I., \&
 Vermeulen, R.~C.\ 2000, \apj, 530, 233


 \bibitem[White et al.(1997)]{whiet97}
 White, R. L., Becker, R. H., Helfand, D. J., \& Gregg, M. D.
 1997, \apj, 475, 479 (FIRST)





 \bibitem[Wrobel(2000)]{wro00}
 Wrobel, J. M. 2000, \apj, 531, 716

 \bibitem[Wrobel, Fassnacht, \& Ho(2001)]{wroet01} 
 Wrobel, J.~M., Fassnacht, C.~D., \& Ho, L.~C.\ 2001, \apjl, 553, L23

 \bibitem[Wrobel \& Heeschen(1984)]{wrohee84}
 Wrobel, J. M., \& Heeschen, D. S. 1984, \apj, 287, 41

 \bibitem[Wrobel \& Heeschen(1991)]{wrohee91}
 Wrobel, J. M., \& Heeschen, D. S. 1991, \aj, 101, 148

 \bibitem[Wrobel et al.(2004)]{wroet04}
 Wrobel, J. M., Machalski, J., \& Condon, J. J., in preparation 

 \bibitem[Wrobel et al.(1996)]{wroet96}
 Wrobel, J. M., Walker, R. C., \& Bridle, A. H.  1996, in Extragalactic 
 radio sources: proc. of the 175th Symposium of the IAU, ed. R. D. Ekers, 
 C. Fanti, \& L. Padrielli (Kluwer Academic Publishers), 131


 \bibitem[Xu et al.(2000)]{xuet00} 
 Xu, C., Baum, S.~A., O'Dea, C.~P., Wrobel, J.~M., \& 
 Condon, J.~J.\ 2000, \aj, 120, 2950 



\bibitem[Zensus(1997)]{zen97}
Zensus, J. A. 1997, \araa, 35, 607

 \bibitem[Zirbel \& Baum(1995)]{zirbau95}
 Zirbel, E. L., \& Baum, S. A. 1995, \apj, 448, 521

\end{thebibliography}
\end{document}